\documentclass{article}

\usepackage[utf8]{inputenc} 
\usepackage[T1]{fontenc} 
\usepackage{hyperref}
\usepackage{url} 
\usepackage{booktabs} 
\usepackage{amsfonts,dsfont,amsmath,amsthm,mathtools} 
\usepackage{nicefrac} 
\usepackage{microtype} 
\usepackage[bottom]{footmisc} 
\usepackage{natbib,authblk} 
\usepackage{float} 
\usepackage{comment} 
\usepackage[para,online,flushleft]{threeparttable} 

\newtheorem{theorem}{Theorem}

\newtheorem{remark}[theorem]{Remark}

\providecommand{\keywords}[1] {\small \textbf{\textit{Keywords---}} #1}

\usepackage{todonotes}

\newcommand{\EE}{\mathbb{E}}
\newcommand{\bx}{\mathbf{x}}
\newcommand{\by}{\mathbf{y}}
\newcommand{\bz}{\mathbf{z}}

\newcommand{\bY}{\mathbf{Y}}
\newcommand{\bZ}{\mathbf{Z}}

\title{Optimal Execution with Quadratic Variation Inventories}

\author[1]{Ren\'e Carmona}
\author[1]{Laura Leal}
\affil[1]{\small Department of Operations Research and Financial Engineering,
  Princeton University, 
  Princeton, NJ 08544,
  
 \texttt{\{rcarmona, lleal\}@princeton.edu}}

\date{}

\begin{document}
\maketitle

\begin{abstract}
The first half of the paper is devoted to description and implementation of statistical tests arguing for the presence of a Brownian component in the inventories and wealth processes of individual traders. We use intra-day data from the Toronto Stock Exchange to provide empirical evidence of this claim. We work with regularly spaced time intervals, as well as with asynchronously observed data. The tests reveal with high significance the presence of a non-zero Brownian motion component. The second half of the paper is concerned with the analysis of trader behaviors throughout the day. We extend the theoretical analysis of an existing optimal execution model to accommodate the presence of It\^o inventory processes, and we compare empirically the optimal behavior of traders in such fitted models, to their actual behavior as inferred from the data.

\keywords{Quantitative Finance, High-Frequency Econometrics;
Quadratic Variation; Presence of Brownian Motion}
\end{abstract}

\pagebreak

\section{Introduction}

High-frequency trading is a computerized practice based on algorithms that allow firms to trade stocks and other financial instruments in microseconds, or even smaller fractions of a second. Over the past two decades, the use of statistical and econometric methods to analyze high-frequency financial data has seen a huge growth. Mathematical finance has long been involved in modeling price processes, which amounts to specifying their dynamics. The key in this modeling is to take into account the randomness of the quantities of interest. If randomness were not involved, we would only need to specify a differential equation concerning the time evolution of the process. 

Accounting for randomness in price processes has evolved from the use of standard It\^o processes to include jumps through the introduction of Poisson and L\'evy processes, to general Markov processes and semi-martingales, and to processes which are  neither Markov nor semi-martingales with the introduction of terms driven by fractional Brownian motions. Meanwhile, the dynamics of the inventory and wealth processes of the trading agents has gone largely ignored. Both have traditionally been modeled as differential functions of time (see \cite{cjp15} for an overview) and up until very recently, as in \cite{kalsi2020optimal,cartea2020optimal}. 

When studying the price process, a significant part of the published literature on high-frequency econometrics focuses on the development of statistical tests for the presence, or absence thereof, of Brownian motion or jump components in price processes. 
Addressing this concern, \cite{aj10} provides two tests for the presence of a continuous part in a semi-martingale model of high-frequency prices. The first test considers the null hypothesis that the Brownian motion is present, while the second one considers the null hypothesis that it is absent and the process is driven by a pure jump process. From both tests, the authors conclude that a continuous component should be included to model the price process. \cite{jkl12} proposes a ratio test for the null hypothesis that the Brownian motion is present based on counting returns smaller than a given threshold. 
\cite{cm11} considers a semi-martingale whose jump component is a Lévy process. They propose two tests: one for the presence of a continuous martingale component in the price process, which allows to discriminate between pure-jump and jump-diffusion models, and another to determine whether the jump component has finite or infinite variation.

The bulk of econometric analyses focuses on price processes. In this paper, we focus instead on the agent's stock inventory and wealth processes. Most of the papers found in the existing literature assume that inventories can be modeled by bounded variations functions, and for practical purposes, even differentiable functions. We follow the preliminary analysis found in the appendix of \cite{CarmonaWebster} and argue with a thorough implementation of statistical tests that both the inventory and the wealth processes of individual traders have a Brownian component. 

For regularly sampled observations, we perform the same test as \cite{aj10}, such as described in section 13.2.1 of \cite{aj14}. The idea is to add a {\em fictitious} Brownian motion to the observed data, coming from a simulated Wiener process, and test whether the integrated volatility is equivalent to that of the fictitious Brownian motion, or if it is larger. For irregularly sampled, or asynchronous, observations, we create an equivalent test for the absence of Brownian motion which takes into account the time irregularity.

Once we show that both the inventory and the wealth of a trader should typically be modeled by including Brownian motion components, we investigate the consequences of such a departure from the main stream literature on the problem of optimal execution. For the sake of definiteness, we choose a popular model in the financial engineering literature, \cite{cj16} to be specific, and we generalize its premises to include Brownian motions components so that the time evolutions of the inventory and the wealth are not given by differentiable functions any longer. We solve this extended model explicitly using the stochastic maximum principle instead of the analytic approach followed in \cite{cj16}.

In order to illustrate the significance of the consequences of our theoretical results, we compare their implementations to the actual inventory and wealth processes of active traders using real trading data from the Toronto Stock Exchange. For that, we simulate Monte Carlo trajectories that include the new proposed component. We do so in two different setups. In the first approach, we keep the price process observed in the data when constructing the simulated trajectories. In the second approach, we simulate the price process by including the market impact that the execution of a large order has on the limit order book according to the model. Moreover, we use these two approaches to understand the behavior of a well-known trader in two different situations: when executing a large order, and when acting as a market maker.

The paper is organized as follows.
In Section \ref{sec:2}, we argue the presence of Brownian motion components in the inventory and wealth processes through the implementation of statistical hypothesis tests. The first set of tests, for regularly spaced time steps, is performed in Section \ref{sec:2.1}. The second set of tests, for asynchronous observations, is described and executed in Section \ref{sec:2.2}. Building on the evidence provided by the statistical tests, Section \ref{sec:3} introduces a natural generalization of an existing model for optimal execution from \cite{cj16}, that takes into account the fact that the inventory process has a Brownian component. We construct an explicit solution in Section \ref{sec:3.1} using the stochastic maximum principle. Two approaches for the practical implementation of the model are proposed in Section \ref{sec:3.2}, and their numerical implementations are detailed in Section \ref{sec:3.3}. Section \ref{sec:4} concludes.

\vskip 6pt\noindent
\emph{Acknowledgement:} The authors were partially supported by the following grants: {\tt NSF DMS-1716673}, {\tt ARO W911NF-17-1-0578}, and {\tt AFOSR FA9550-19-1-0291}.

\section{Statistical Tests for the Presence of a Brownian Component}
\label{sec:2}
We are investigating whether the path of a trader's inventory $Q$ and their wealth process $X$ should be modelled with a diffusion component, instead of the mere time-dependent drift that is usually considered in the literature (see \cite{cjp15} for a wide range of different configurations of the optimal execution problem). This section is devoted to the implementation of statistical tests showing that if we assume that the inventory and the wealth processes of a trader are It\^o processes, the absence of Brownian components should be rejected in essentially all cases. Obviously, one can easily imagine a trader placing orders of a fixed size at regularly spaced times, forcing their inventory to be linear in time, hence without a Brownian component. However, this concocted example does not occur in practice as traders do not want to be detected and subsequently frontrun. 

\subsection{The Case of Regular Observations}
\label{sec:2.1}
In order to present the theoretical results used to derive our statistical tests, we use the notation $Q$ for a generic continuous It\^o process, and urge the reader to keep in mind that the same test will also be applied to the wealth process $X$. 

The underlying process $Q$ is assumed to be a one-dimensional continuous semi-martingale, defined on a filtered space $(\Omega,\mathcal{F},(\mathcal{F})_{t\geq 0},\mathbb{P})$ of the form:
\begin{equation}
    Q_t = Q_0 + \int_0^t b_s ds + \int_0^t \sigma_s dW_s,
    \label{eq:8}
\end{equation}

\noindent where $b_t$ is an $\mathbb{R}$-valued progressively measurable process, $\sigma_t$ is a predictable $\mathbb{R}$-valued process, $W$ is a one-dimensional Brownian motion.
The characteristics $(B,C)$ are defined as:
\begin{equation}
    B_t = \int_0^t b_s ds, \hspace{5mm} C_t = \int_0^t c_s ds,
    \label{eq:9}
\end{equation}

\noindent so $B_t$ is a predictable process of finite variation, and $C_t$ is the quadratic variation of the continuous local martingale part, with $c=\sigma^2$.

A remarkable result in \cite{cm11} allows us to separate the null hypothesis $H_0$ that the Brownian motion is not present, from the alternative hypothesis indicating the Brownian-driven term is actually present:

\begin{equation}
    \begin{cases}
    H_0: \Omega_T^{(noW)}=\{C_T=0\},\\
    H_1: \Omega_T^{(W)}=\{C_T>0\}.
    \end{cases}
\end{equation}

The null hypothesis here represents the event on which the quadratic variation of the continuous local martingale part, considered over the full path of the process, is equal to zero. This means that the diffusion component is not present in the process, and therefore we should not use a Brownian motion to model this process. On the other hand, if we reject the null hypothesis $H_0$, we conclude that the event $\Omega_T^{(W)}=\{C_T>0\}$ is likely the true scenario for the observation, and that the Brownian motion term should be used when modelling the inventory process $Q$ of the trading agent (or the wealth process $X$ of the agent). 

For this first test, we consider the observations to be regularly spaced over a fixed time interval $[0,T]$. Hence, they occur at times $i\Delta_n$, for $i\in{0,1,\dots}$ where $\Delta_n \rightarrow 0$.

\paragraph{Assumption (H1)} We assume that the process $b$ is locally bounded and that $\sigma$ is c\`adl\`ag (right continuous with left limits).

\vspace{5mm}

For the test introduced by \cite{cm11} and described in \cite{aj14}, since the null hypothesis is that there is no Brownian component, we add a fictitious Brownian motion $W'$ to the original process observed in the data, and generate the increments $\Delta_i^n Q'$ of an artificial process $Q'=Q+\sigma' W'$.

\begin{equation}
    \Delta_i^n Q' = \Delta_i^n Q + \sigma ' \Delta_i^n W',
\end{equation}

\noindent where $\Delta_i^n W'$ is independent of the data, and we can assume $W'$ to be defined on $(\Omega,\mathcal{F},\mathbb{P})$ and adapted to the filtration $\mathbb{F}=(\mathcal{F}_t)_{t\ge 0}$. In this test, we consider $\sigma '>0$ to be a constant, and take $c'=\sigma^2$. 

Notice that adding the fictitious Brownian motion to the inventory process has the effect of changing the second characteristic of the process $Q$ as defined in \eqref{eq:9} to $C_t'=C_t + tc'$ for the process $Q'$. 

If we take the second characteristic at time $T$, we can test the following combination of null and alternative hypothesis:
\begin{equation}
    \begin{cases}
      \Omega_T^{(noW)} = \{C_T' = c'T\},\\
      \Omega_T^{(W)} = \{C_T' > c'T\}.
    \end{cases}
    \label{eq:hyp_test}
\end{equation}

Notice also that, \textit{under the null hypothesis}, the second characteristic of the original process $Q$ is assumed to be equal to zero. From the second equation in \eqref{eq:9}, using $c=\sigma^2$, this is equivalent to saying that $\sigma = 0$. Thus, we should not use the Brownian component in equation \eqref{eq:8} to model the process $Q$. We will see that rejecting the null hypothesis will support our claim that the \textit{inventory process $Q$ should be modelled using a Brownian-driven component.}

In order to show that, we first specify a truncation level $u_n \asymp \Delta_n^{\bar{\omega}}$, for some $\bar{\omega}\in \big(0,\frac{1}{2}\big)$.\footnote{Notation: $u_n \asymp v_n$ means that $\frac{1}{A} \leq \frac{u_n}{v_n} \leq A$, for a constant $A\in(1,\infty)$.} We take 
\begin{equation}
    u_n = \gamma \eta^{1/2} \Delta_n^{1/2},
    \label{eq:truncation_un}
\end{equation}

\noindent i.e. we take the truncation level to be a number $\gamma$ of standard deviations of the Brownian part, where $\eta^{1/2}$ is the long term volatility, and $\Delta_n$ is the length of the regular time interval. The motivation behind truncating the process is to eliminate possible large jumps in the process while keeping the increments from the continuous part of the process, if there are any. Hence, we would like to pick a truncation level $u_n$ as small as possible such that not too many Brownian increments are discarded.

Once the truncation level is chosen, we define the truncated realized volatility of the process $Q'$ as:
\begin{equation}
    \hat{C}^{Q'}(\Delta_n,u_n)_T = \sum_{i=1}^{[T/\Delta_n]} (\Delta_i^n Q')^2 \mathds{1}_{\{|\Delta_i^n Q'|\leq u_n\}},
\end{equation}
\noindent which is the realized volatility given that the process is truncated to exclude large jumps. Its associated quarticity estimator is defined as:
\begin{equation}
    \hat{B}^{Q'}(4,\Delta_n,u_n)_T = \sum_{i=1}^{[T/\Delta_n]} (\Delta_i^n Q')^4 \mathds{1}_{\{|\Delta_i^n Q'|\leq u_n\}}.
    \label{eq:quarticity}
\end{equation}

\noindent where $\hat{B}^{Q'}(4,\Delta_n,u_n)_T \overset{p}{\to} 3 \int_0^T(c_s+c')^2 ds$. The quarticity estimator is necessary in order to use feasible central limit theorems for volatility estimators, and hence feasible confidence intervals. There are other methods to estimate quarticity (see \cite{bns03, m12, ads11}), but the estimator in \eqref{eq:quarticity} is sufficient for our purposes.

Under Assumption (H1), we can define the test statistics, by Theorem 6.10 in \cite{aj14}: 
\begin{equation}
    \sqrt{\frac{3}{2}} \frac{\hat{C}^{Q'}(\Delta_n,u_n)_T  - C_T'}{\sqrt{\hat{B}^{Q'}(4,\Delta_n,u_n)_T}} 
\end{equation}

\noindent whose convergence to $\mathcal{N}(0,1)$ is stable, with the Gaussian limit independent of the filtration $\mathbb{F}$.

Since we are performing a one-sided test, we choose the critical region to be:
\begin{equation}
    C_n = \bigg\{ \hat{C}^{Q'}(\Delta_n,u_n)>c'T + \frac{z_{\alpha}'}{\sqrt{3}}\sqrt{2 \hat{B}^{Q'}(4,\Delta_n,u_n)_T }\bigg\}.
\end{equation}
as rejection region for the null hypothesis. If we are able to reject the null hypothesis for the absence of Brownian motion in the inventory, and later in the wealth, then we can confidently assert that both these processes should be modelled in the way we propose. In the two paragraphs below, we show the result from these hypothesis tests on data regularly sampled in five minute intervals.

\subsubsection{Implementation with Real Trader Inventory Processes}
\label{sec:2.1.1}
\begin{figure}[H]
\centering
\begin{minipage}{.45\linewidth}
  \includegraphics[width=\linewidth]{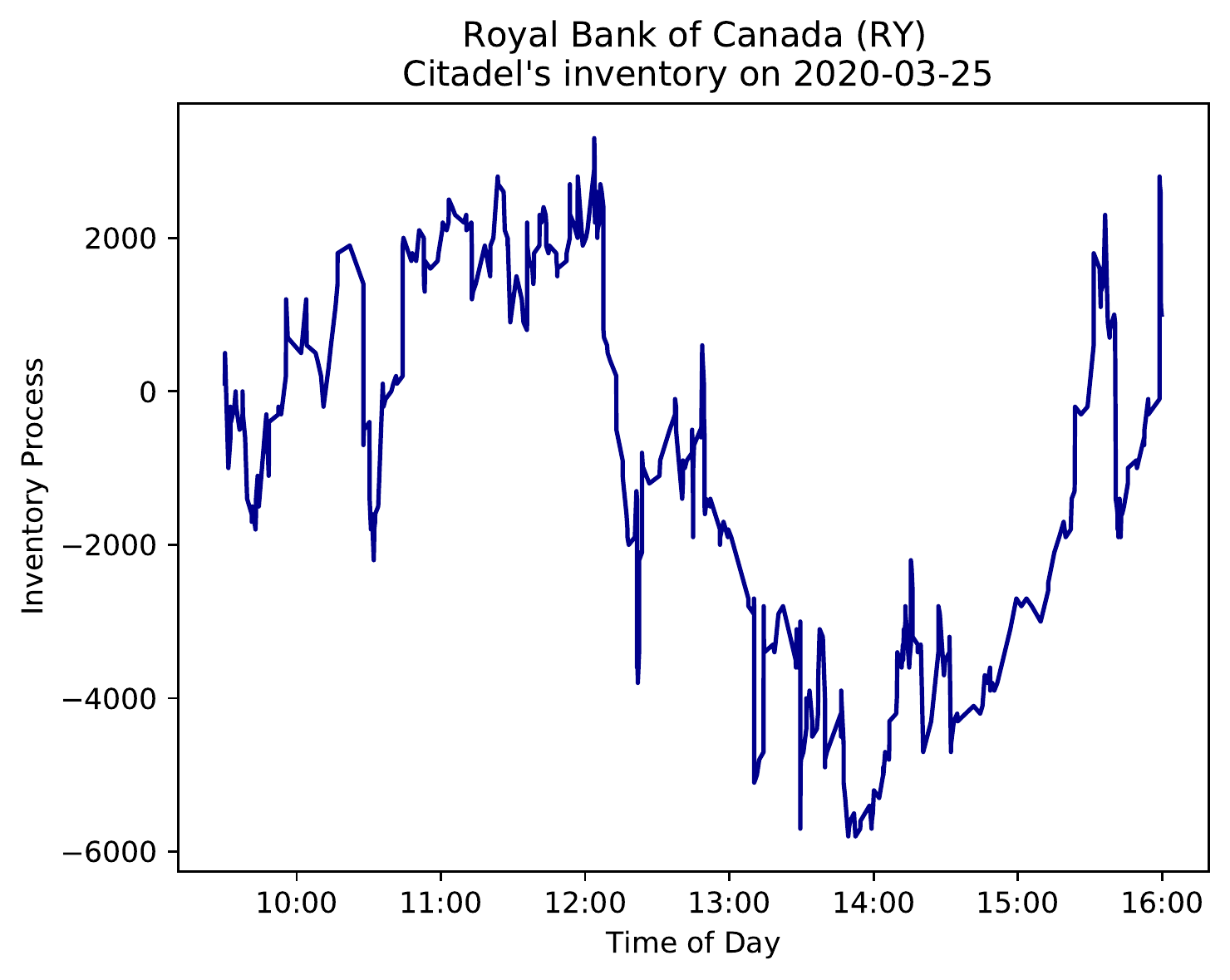}
\end{minipage}
\hspace{10mm}
\begin{minipage}{.45\linewidth}
  \includegraphics[width=\linewidth]{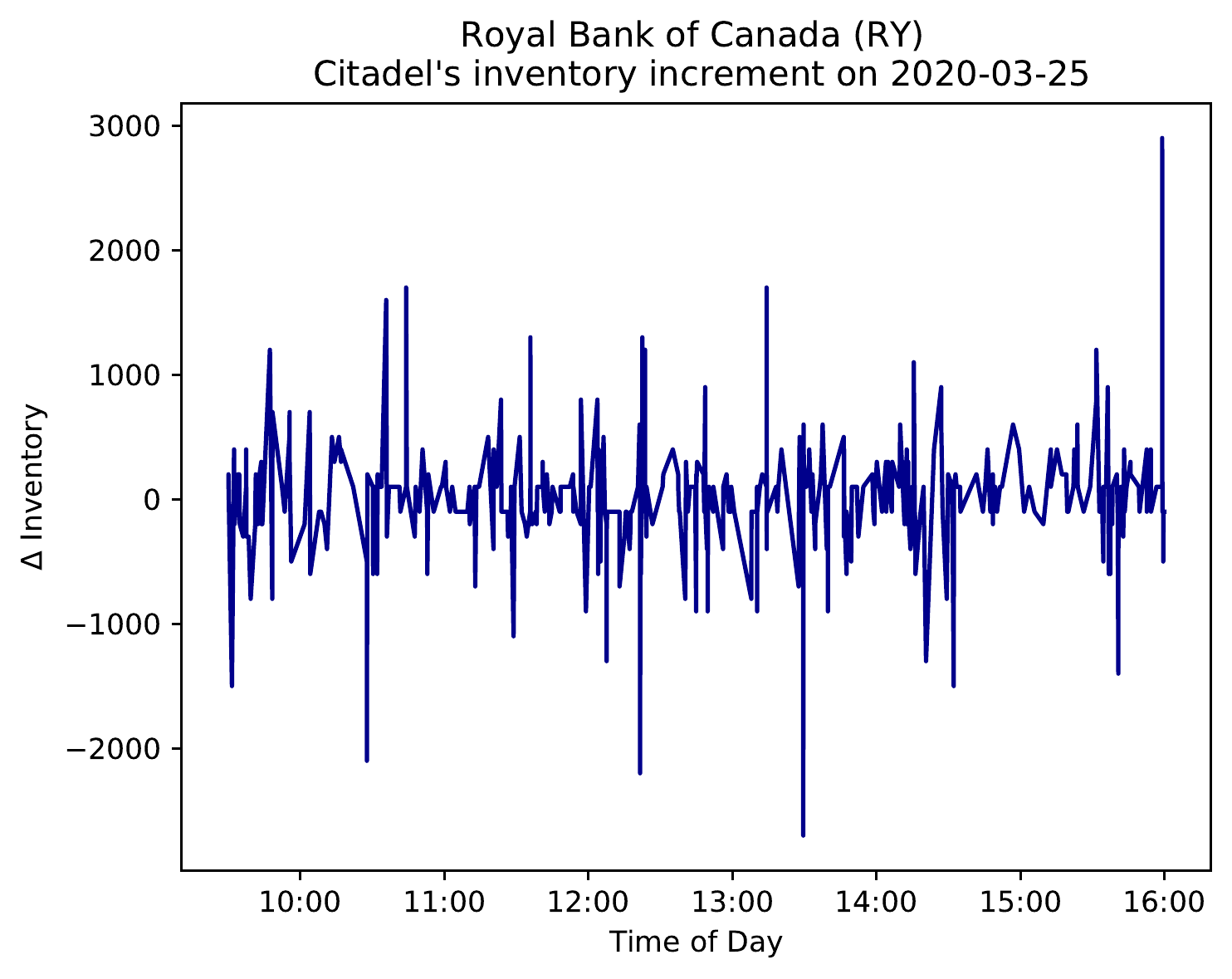}
\end{minipage}
\caption{Inventory $Q$ process path (left) and asynchronous increments (right), on March $25^{th}$, 2020}
\label{fig:1}
\end{figure}

\begin{figure}[H]
\centering
\begin{minipage}{.45\linewidth}
  \includegraphics[width=\linewidth]{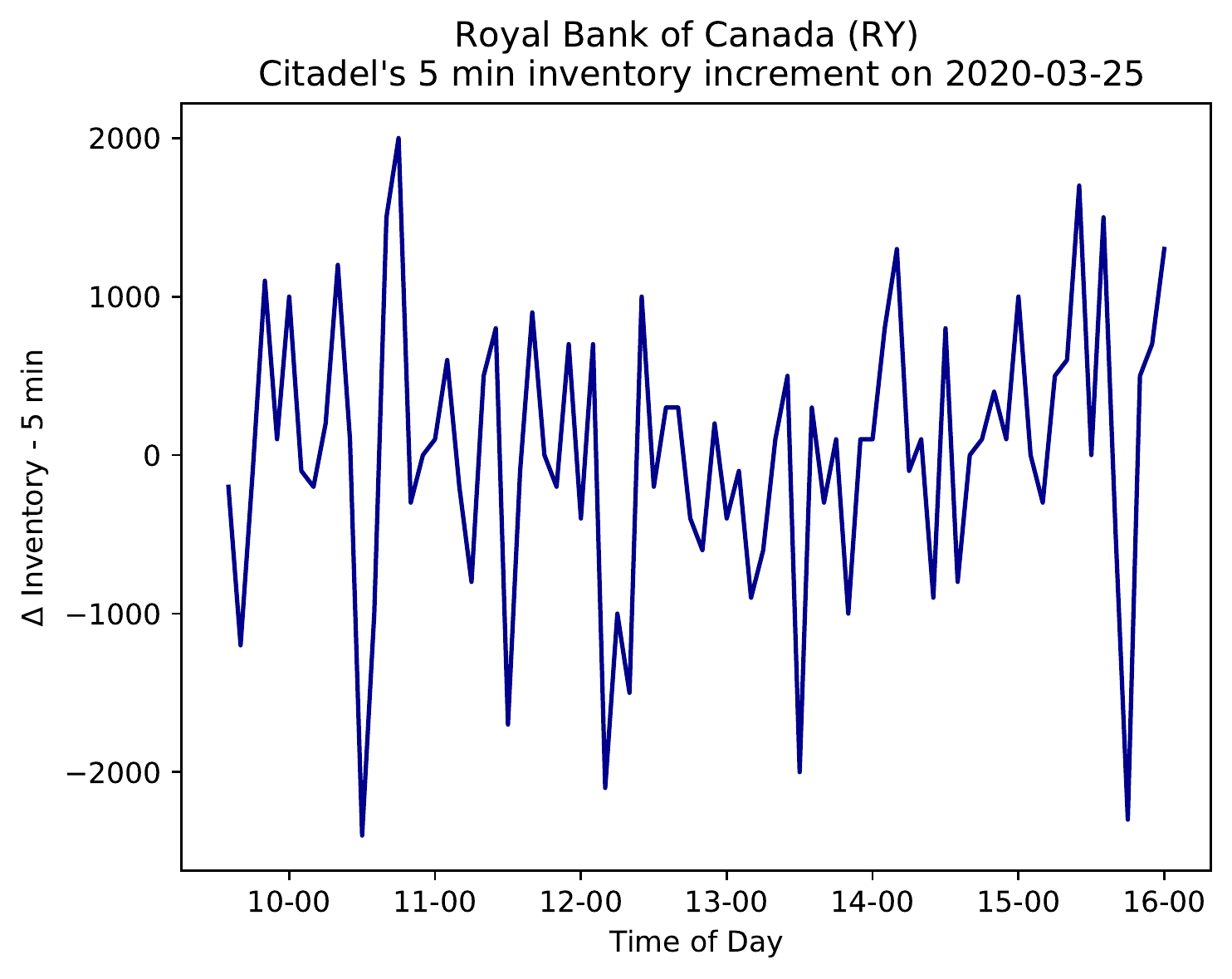}
\end{minipage}
\hspace{10mm}
\caption{Inventory $Q$ process increments on regular five-minute increments, on March $25^{th}$, 2020}
\label{fig:2}
\end{figure}

 We use data for the stock of Royal Bank of Canada (RY) and select an active trader, Citadel Securities, to test whether their inventory process should be modelled using a Brownian motion component. The plot on the left of Figure \ref{fig:1} shows the realization of the inventory process $Q$ on March 25, 2020, while the one on the right shows its associated asynchronous increments. Notice that the right plot has two large negative observations that at first glance do not seem to correspond to the left plot, and the left plot has a large negative slope that does not seem to be reflected in the right plot. In the first case, the increments are only individual observations, and we do see drops in the left plot with the magnitude of the increments in the right plot. In the second case, while there are no large observations, there are several trades in a sequence, likely pertaining to one large meta-order. We can see more clearly in Figure \ref{fig:2} how these small negative trades add up once we aggregate them over five minute intervals. 

In this section, we re-sample the data to regularly spaced time increments $\Delta_n$ of five-minute intervals. The increments resulting from this resampling to regularly spaced time intervals can be seen in Figure \ref{fig:2}.

When performing the test for the data in five-minute bins, we find that it is sensitive to two important parameters: the standard deviation $\sigma '$ associated to the Brownian fictitious motion $W'$, and the standard deviation $\gamma$ defining the truncation level $u_n$.

\begin{figure}[H]
\centering
\begin{minipage}{.45\linewidth}
  \includegraphics[width=\linewidth]{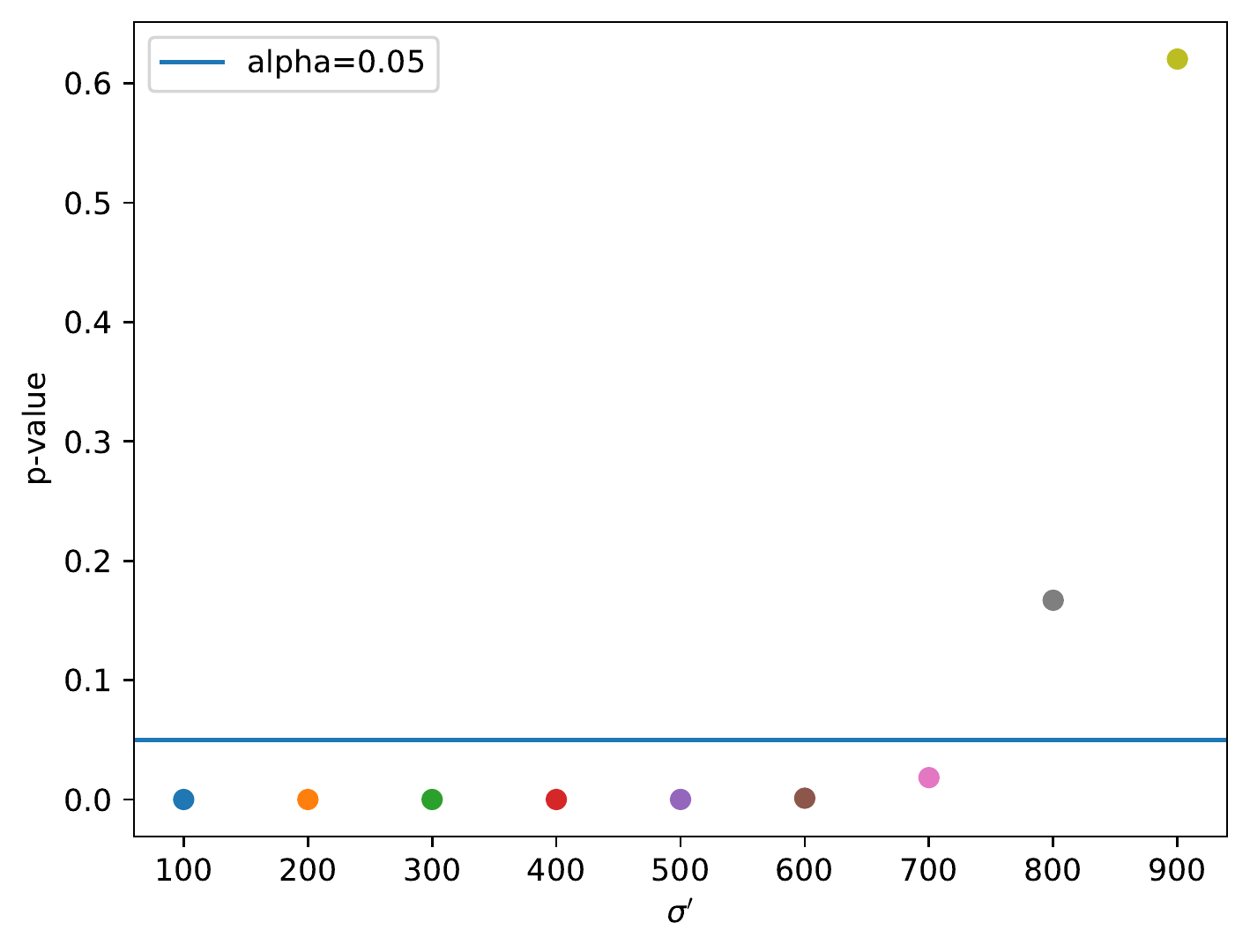}
\end{minipage}
\hspace{10mm}
\begin{minipage}{.45\linewidth}
  \includegraphics[width=\linewidth]{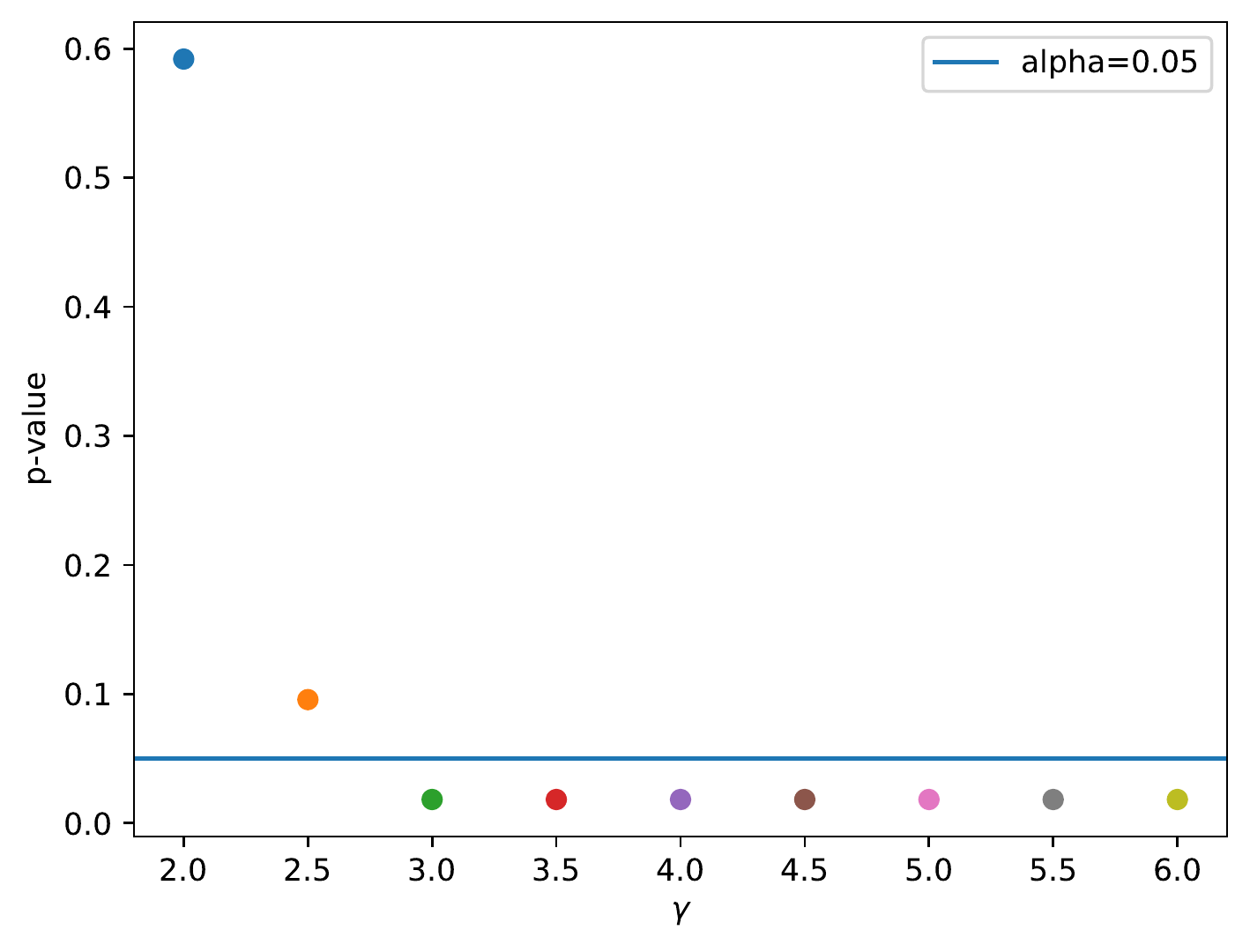}
\end{minipage}
\caption{P-value of the test: the influence of $\sigma'$ and $\gamma$}
\label{fig:3}
\end{figure}

In Figure \ref{fig:3}, the horizontal line shows a level of confidence $\alpha=5\%$ chosen for the sake of definiteness. The left plot uses a fixed $\gamma=3$ while doing a sensitivity analysis for $\sigma '$. For values $\sigma '\leq700$, we consistently reject the null hypothesis and conclude that we should include the Brownian motion component when modelling the inventory.  For larger values of $\sigma '$, we cannot reject the null hypothesis. Indeed, if $\sigma '$ is very large, it becomes very hard to meet the alternative hypothesis of the test, namely that the second characteristic of the process is greater than the quadratic variation of the continuous local martingale at the terminal time $T$. Clearly, decreasing the parameter $\sigma '$ increases the power of the test. Thus, we argue that we should take $c'T$ in \eqref{eq:hyp_test} to be small relative to the quadratic variation of the inventory process $Q$. 

The right plot of Figure \ref{fig:3} fixes $\sigma '= 700$, and allows $\gamma$ to vary. We find that, for any reasonable level of the standard deviation parameter $\gamma$ defining the truncation level, we can reject the null hypothesis that the Brownian component is absent. Note that we consider $\gamma=3$ to be a reasonable level for cutting out jumps from the process because it is the usual value used to search for extreme observations in the normal distribution. However, for any stricter value, $\gamma>3$, the test will result in the same conclusion.

To demonstrate that the above conclusion is not merely anecdotal, but ubiquitous, we perform the test for absence of Brownian motion on regularly timed observations for ten stocks (ABX, AC, BMO, BNS, ENB, MFC, RCI.B, RY, SU, and TD), for all traders listed in the Toronto Stock Exchange (TSX), for all trading days between January $2^{nd}$, 2010 to October $26^{th}$, 2020. We find that a Brownian component is relevant to the inventory process on the vast majority of days, as can be noted from Table \ref{tab:inventory_BM_reg}. In order to be included in the computation, the only restriction we make is that on a given day the trader  traded on average at least once per minute, but we make no assumption on the distribution of trades throughout the day. 

\vskip 12pt

\begin{table}[H]
\centering
\begin{tabular}{lrrrrrrrrrr}
\hline
   &    ABX &     AC &    BMO &    BNS &    ENB &    MFC &  RCI.B &     RY &     SU &     TD \\
\hline
Trader $\#2$  &   99.5 &   99.8 &   99.5 &   99.6 &   99.6 &   99.7 &   99.6 &   99.6 &   99.6 &   99.4 \\
Trader $\#5$  &  100.0 &  100.0 &  100.0 &  100.0 &  100.0 &  100.0 &  100.0 &  100.0 &  100.0 &  100.0 \\
Trader $\#7$  &   99.7 &   99.9 &   99.8 &   99.8 &   99.4 &   99.7 &   99.8 &   99.6 &   99.7 &   99.8 \\
Trader $\#14$ &   98.2 &   98.7 &   99.7 &   98.8 &   98.9 &   96.9 &  100.0 &   99.4 &   98.8 &   98.4 \\
Trader $\#15$ &   99.7 &   99.5 &   99.6 &   99.5 &   99.2 &   99.3 &   99.4 &   99.0 &   99.5 &   99.4 \\
Trader $\#39$ &   99.7 &  100.0 &   99.5 &   99.5 &   99.5 &   99.6 &   99.2 &   99.3 &   99.5 &   99.7 \\
Trader $\#53$ &   99.3 &   99.6 &   99.2 &   99.3 &   99.5 &   99.3 &   99.5 &   99.5 &   99.4 &   99.5 \\
Trader $\#65$ &   98.8 &   98.3 &   99.3 &   99.6 &   99.4 &   99.2 &   98.7 &   98.9 &   98.4 &   99.3 \\
\hline
\end{tabular}
\caption{Percentage of days for which there is Brownian component in regularly sampled observations for the inventory data of active traders.}
\label{tab:inventory_BM_reg}
\end{table}

Therefore, given that the data has been regularly sampled, we have argued that the \textit{trader's inventory process $Q$ should be modelled including a Brownian motion component.}

\subsubsection{Implementation with Real Trader Wealth Processes}
\label{sec:2.1.2}
\begin{figure}[H]
\centering
\begin{minipage}{.45\linewidth}
  \includegraphics[width=\linewidth]{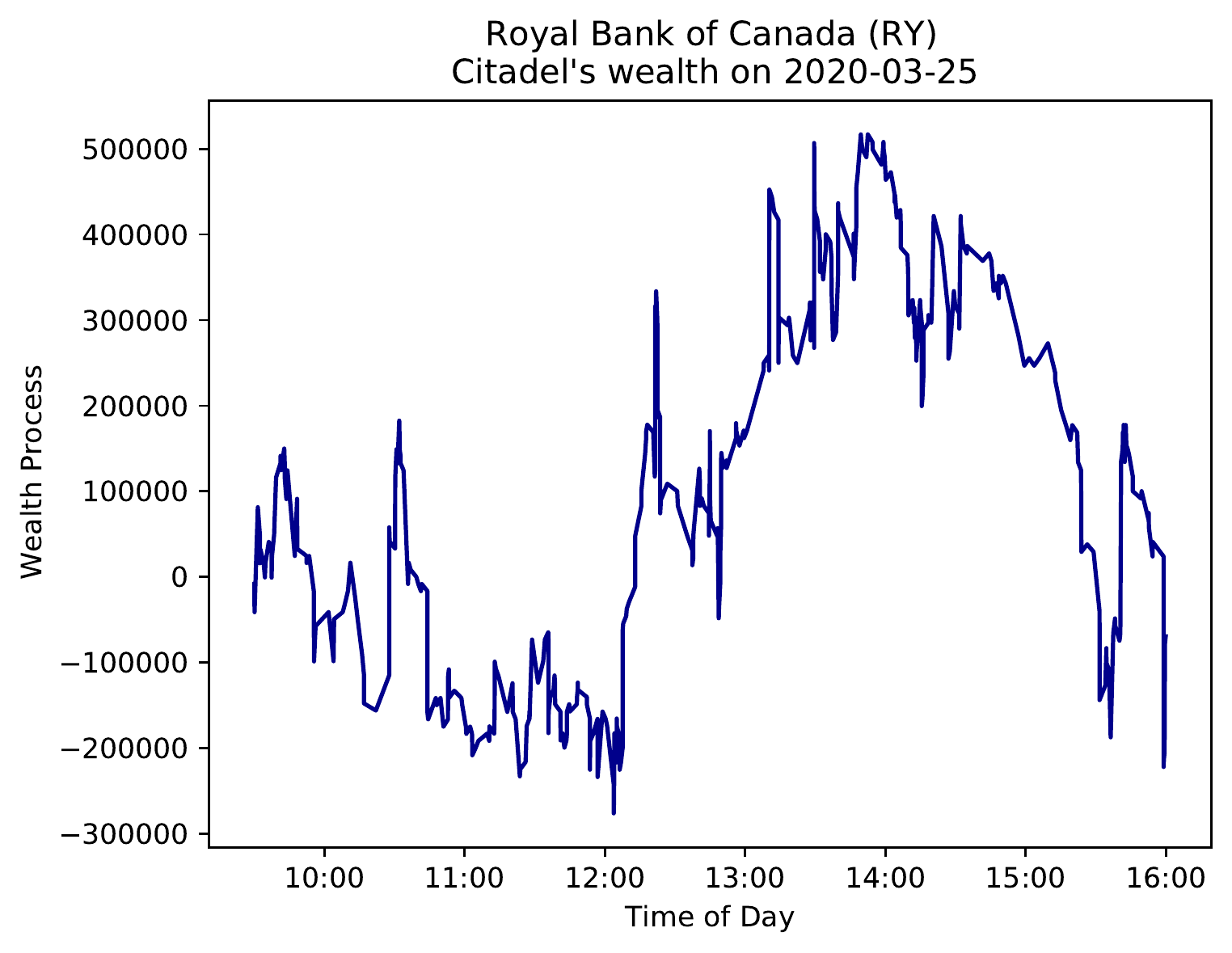}
\end{minipage}
\hspace{10mm}
\begin{minipage}{.45\linewidth}
  \includegraphics[width=\linewidth]{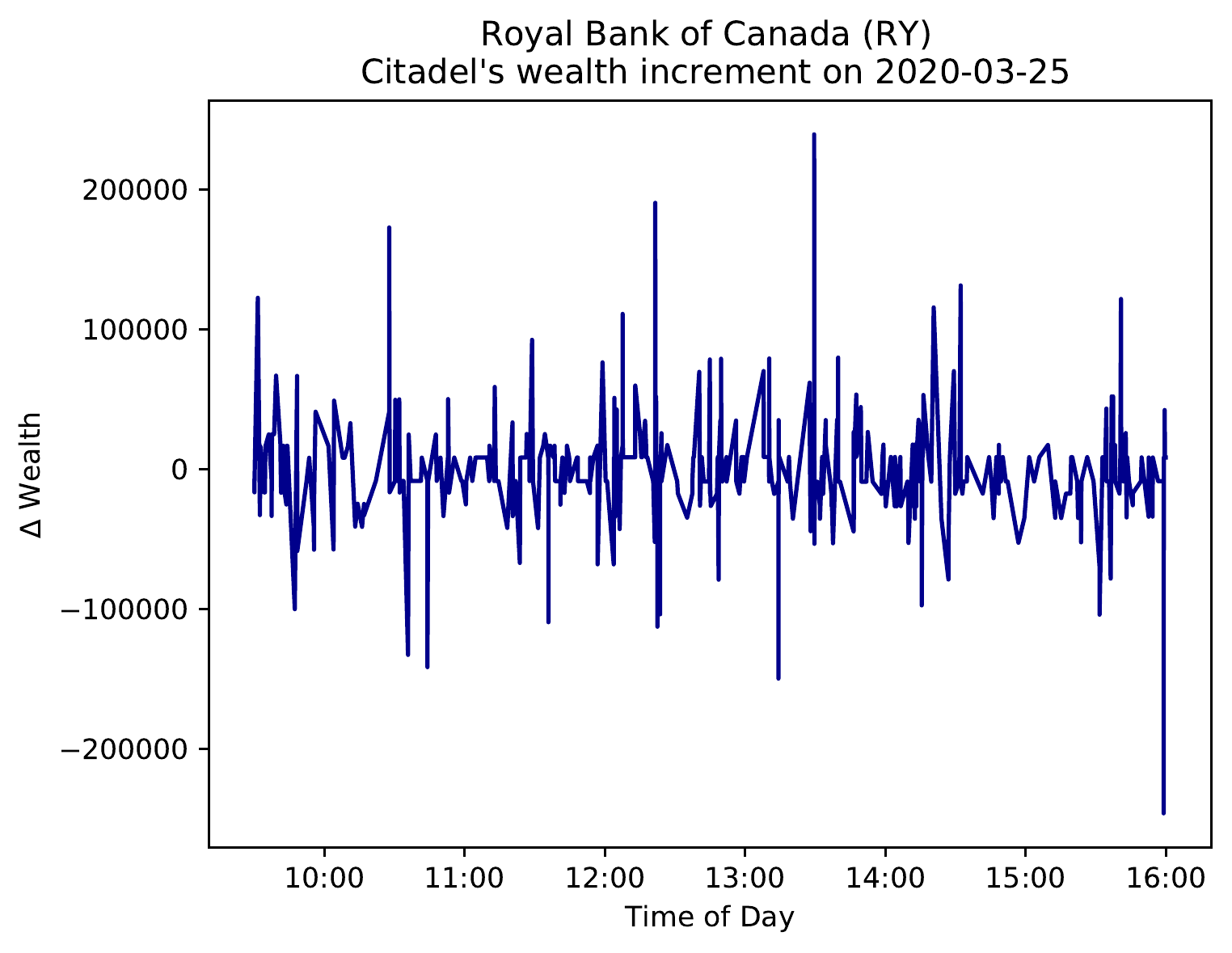}
\end{minipage}
\caption{Wealth $X$ process path (left) and asynchronous increments (right), on March $25^{th}$, 2020}
\label{fig:4}
\end{figure}

Again, we take data for the stock of Royal Bank of Canada (RY) and select Citadel Securities as the trader. Now, we want to test whether their wealth process should be modelled using a Brownian motion component. The plot on the left of Figure \ref{fig:4} shows the realization of the wealth process $X$, while the one on the right shows its associated asynchronous increments. We note that large increments in the wealth process are often due to small, yet consistent, trading in a small amount of time. Therefore, large increments on the left plot are not always seen as large increments on the right plot, but as a sequence of very small ones.

\begin{figure}[H]
\centering
\begin{minipage}{.45\linewidth}
  \includegraphics[width=\linewidth]{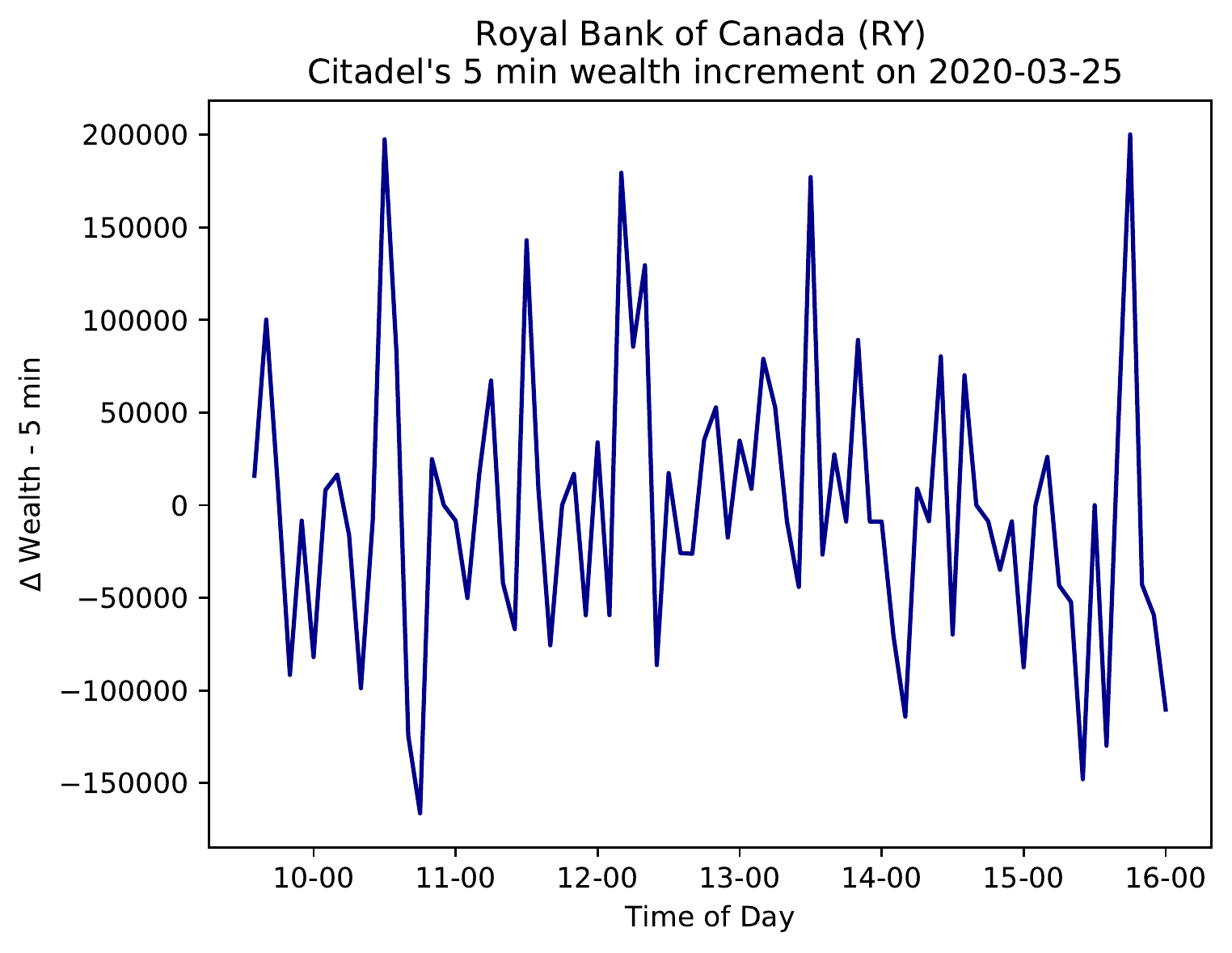}
\end{minipage}
\hspace{10mm}
\caption{Wealth $X$ process increments on regular five-minute increments, on March $25^{th}$, 2020}
\label{fig:5}
\end{figure}

As in the case of the inventory $Q$, the test for the wealth process in this section is conducted for regularly time stamps spaced by five-minute. The wealth increments resulting from this re-sampling can be seen in Figure \ref{fig:5}.

As it was the case for the inventory process, when performing the test for the data in five-minute bins, we find that it is sensitive to two important parameters: the standard deviation $\sigma '$ associated to the Brownian fictitious motion $W'$, and the standard deviation $\gamma$ defining the truncation level $u_n$.

\begin{figure}[H]
\centering
\begin{minipage}{.45\linewidth}
  \includegraphics[width=\linewidth]{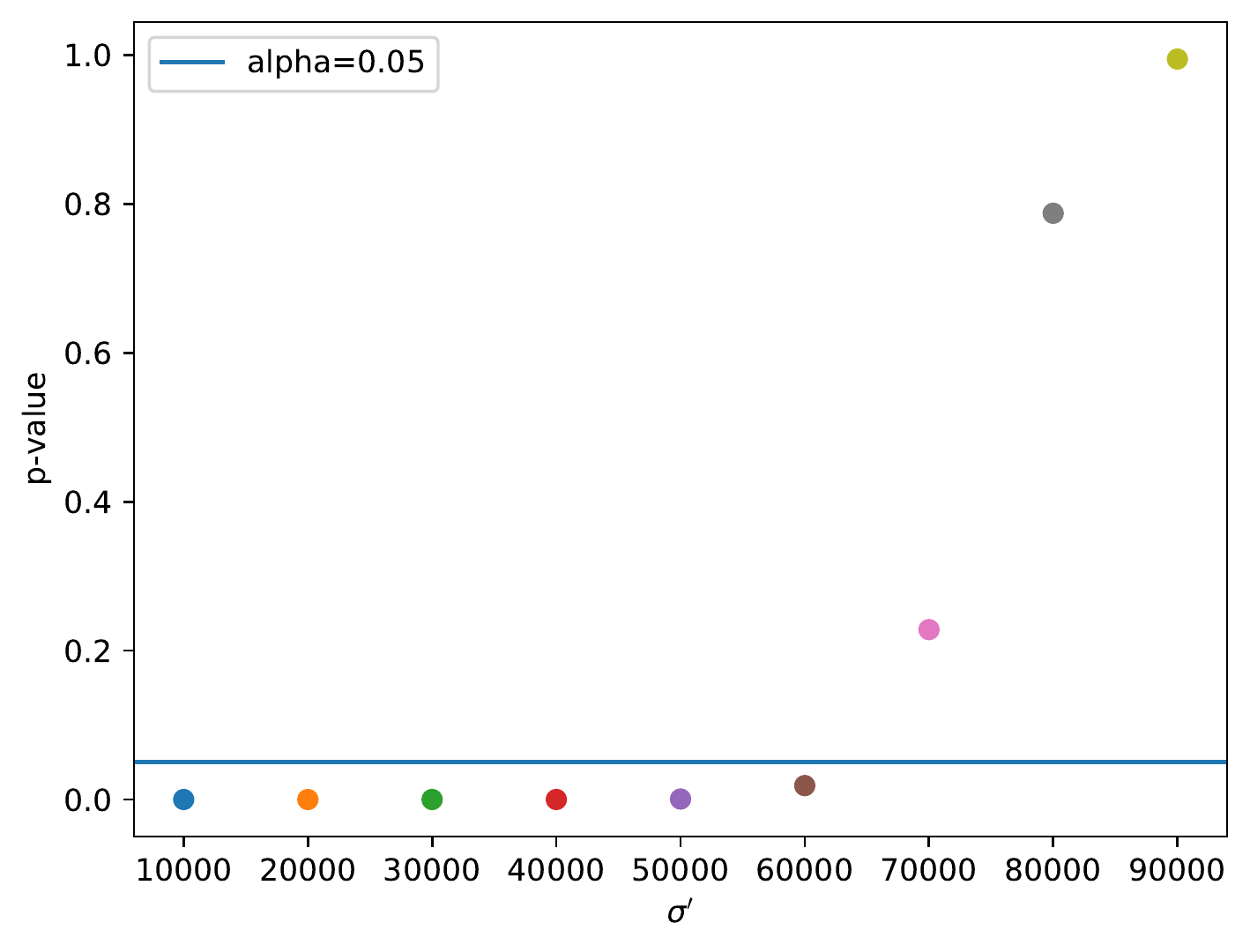}
\end{minipage}
\hspace{10mm}
\begin{minipage}{.45\linewidth}
  \includegraphics[width=\linewidth]{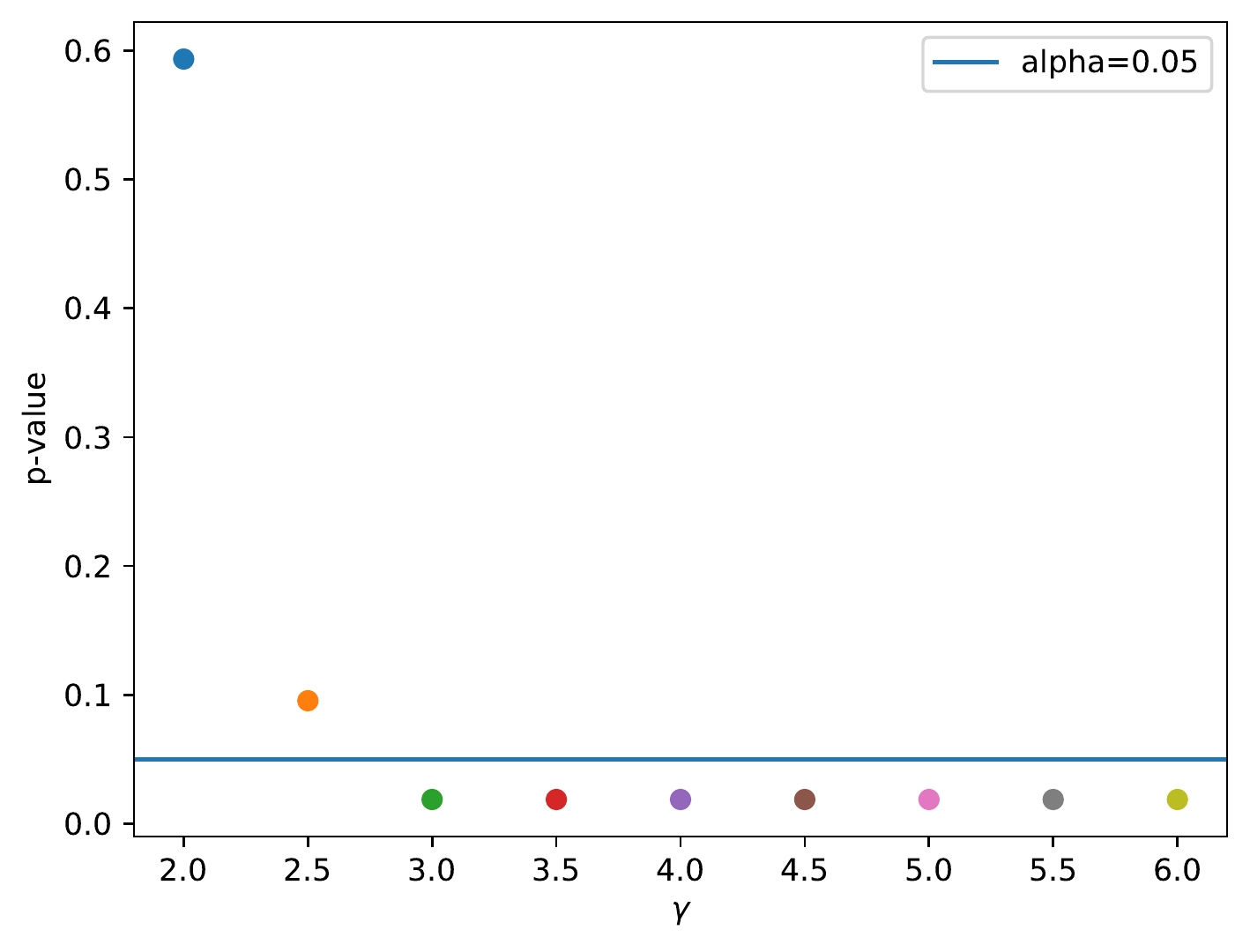}
\end{minipage}
\caption{P-value of the test: the influence of $\sigma'$ and $\gamma$}
\label{fig:6}
\end{figure}

In Figure \ref{fig:6}, the left plot uses the same fixed $\gamma=3$ while doing a sensitivity analysis for $\sigma '$. For values $\sigma '\leq 60000$, we consistently reject the null hypothesis and conclude that we should include the Brownian motion component when modelling the wealth. For larger values of $\sigma '$, we cannot reject the null hypothesis. Note that the threshold for $\sigma '$ in the case of the wealth differs from the inventory case. That is due to the natural differences in the order of magnitude of the two quantities, which is embedded in the price process. For the case of the stock for the test shown in Figure \ref{fig:6}, the price is in the order of \$100.00, hence the wealth is adjusted accordingly. 

The right plot of Figure \ref{fig:6} fixes $\sigma '=60000$, and allows $\gamma$ to vary. For any reasonable level of the parameter $\gamma$ ($\gamma\geq 3$) defining the truncation level, we can reject the null hypothesis that the Brownian component is absent.

As we did for the inventory, we perform the test for absence of Brownian motion on regularly timed observations for the same ten stocks, for all active traders listed on TSX, for all trading days between January $2^{nd}$, 2010 to October $26^{th}$, 2020. We find that a Brownian component is relevant to the wealth process on the vast majority of days, as can be noted from Table \ref{tab:wealth_BM_reg}. 

\vskip 12pt

\begin{table}[H]
\centering
\begin{tabular}{lrrrrrrrrrr}
\hline
   &    ABX &     AC &    BMO &    BNS &    ENB &    MFC &  RCI.B &     RY &     SU &     TD \\
\hline
Trader $\#2$  &   99.5 &   99.9 &   99.5 &   99.6 &   99.6 &   99.7 &   99.6 &   99.6 &   99.6 &   99.4 \\
Trader $\#5$  &  100.0 &  100.0 &  100.0 &  100.0 &  100.0 &  100.0 &  100.0 &  100.0 &  100.0 &  100.0 \\
Trader $\#7$  &   99.7 &   99.9 &   99.8 &   99.8 &   99.4 &   99.7 &   99.8 &   99.6 &   99.7 &   99.8 \\
Trader $\#14$ &   98.2 &   98.7 &   99.7 &   98.8 &   98.9 &   96.9 &   99.7 &   99.4 &   98.6 &   98.4 \\
Trader $\#15$ &   99.7 &   99.5 &   99.6 &   99.5 &   99.2 &   99.2 &   99.4 &   99.0 &   99.5 &   99.4 \\
Trader $\#39$ &   99.7 &   99.9 &   99.5 &   99.5 &   99.5 &   99.6 &   99.2 &   99.3 &   99.5 &   99.7 \\
Trader $\#53$ &   99.2 &   99.6 &   99.2 &   99.3 &   99.5 &   99.3 &   99.5 &   99.6 &   99.5 &   99.5 \\
Trader $\#65$ &   98.8 &   98.5 &   99.3 &   99.6 &   99.4 &   99.3 &   98.7 &   98.9 &   98.4 &   99.3 \\
\hline
\end{tabular}
\caption{Percentage of days for which there is Brownian component in regularly sampled observations for the wealth data, active traders.}
\label{tab:wealth_BM_reg}
\end{table}

Hence, for regularly sampled data, we have argued that the \textit{trader's wealth process $X$ should be modelled including a Brownian motion component}, just like we had concluded for their inventory process.

\vskip 4pt
Next, we test whether a Brownian component should also be included when modelling asynchronous data.

\subsection{The Case of Asynchronous Observations}
\label{sec:2.2}

Transaction prices are recorded at the times of the trades. Hence, they are irregularly spaced in time, or asynchronous. We observe transactions, and thus changes to the inventory of the agent, at times $S_i$, where $S_i<S_{i+1}$. Here, we do not assume any special property of the sequence of observation times, we just assume that they are defined on the original space $\Omega$, and are measurable with respect to a $\sigma$-field which may be larger than $\mathcal{F}$. As in \cite{aj14}, we are open to the possibility of the frequency of observations increasing to infinity, as we would like to make use of asymptotic results. However, on the empirical test, we can observe $S_i$ directly from the data. 

Let $(S(n,i):i>0)$ be strictly increasing observation times, at stage $n$. The stage $n$ is first theoretically defined in \cite{hayashi2011irregular}. From an empirical perspective, it stands for having many observations, which in the case of trading is equivalent to restricting the numerical analysis to active traders. We have $1 \leq i \leq n$, and need $n\rightarrow\infty$ in order to derive asymptotic results. Assume $S(n,0)=0$, which is to say that the trade clock starts at zero. We can define the observation time difference as 
\begin{equation}
    \Delta(n,i) = S(n,i)-S(n,i-1),
\end{equation}

\noindent the number of observation times within the interval $(0,t]$ as $N_t^n=\sup_i \{S(n,i)\leq t\}$, and the ``mesh'' up to time $t$ as $\pi_t^n=\sup_{i=1,\dots,N_t^n +1} \Delta(n,i)$. Moreover, given a fixed time horizon $T$, we assume that $\lim_{i\rightarrow \infty}S(n,i)=\infty, \text{ for all } n, \text{ and } \pi_T^n \overset{\mathbb{P}}{\to} 0, \text{ as } n\rightarrow \infty$. These assumptions allow us to obtain asymptotic results, as the mesh goes to $0$ as $n\rightarrow \infty$, while making sure that there is a finite number of observations at each stage $n$. 

As before, the process $Q$ is assumed to be a one-dimensional continuous Itô semi-martingale of the form
\begin{equation}
    Q_t = Q_0 + \int_0^t b_s ds + \int_0^t \sigma_s dW_s,
    \label{eq:18}
\end{equation}

\noindent defined on a filtered space $(\Omega,\mathcal{F},(\mathcal{F}_t)_{t\geq 0},\mathbb{P})$, and to satisfy the following assumption: 

\paragraph{Assumption (H2)} The process $Q$ is continuous, $b$ is locally bounded, and $\sigma$ is c\`adl\`ag.

\vspace{5mm}
The asynchronous increments of the inventory process $Q$ are then written as 
\begin{equation}
    \Delta_i^n Q = Q_{S(n,i)} - Q_{S(n,i-1)}.
\end{equation}

In order to perform an analogous test to that described in \eqref{eq:hyp_test}, we add a fictitious Brownian motion $W'$ to the original process observed in the data, and generate the increments $\Delta_i^n Q'$ of an artificial process $Q'=Q+\sigma' W'$.
\begin{align}
    \Delta_i^n Q' &= \Delta_i^n Q + \sigma ' \Delta_i^n W',\\
    & = \Delta_i^n Q + \sqrt{c'(S_i - S_{i-1})} \epsilon_i,
\end{align}

\noindent where $\Delta_i^n W'$ is independent of the data, and we can assume $W'$ to be defined on $(\Omega,\mathcal{F},\mathbb{P})$ and adapted to the filtration $\mathbb{F}=(\mathcal{F}_t)_{t\ge 0}$. In the asynchronous version of $\Delta_i^n W'$, the time increment is taken to be the difference between two observation times. The $\epsilon_i$ are i.i.d $\mathcal{N}(0,1)$.  

Then we test the null hypothesis that the cumulative variance is c’T against the alternative that it is greater than c’T, just like in \eqref{eq:hyp_test}. This corresponds to the null that there is no Brownian motion term against the presence of a non-trivial Brownian motion term. Again in the test for the asynchronous case, we consider $\sigma'>0$ to be a constant, and take $c'=\sigma'^2$. 

In order to set the stage to use Theorem 9.3 in \cite{aj14}, let $m_r = \mathbb{E}(|\Phi|^r) = 2^{r/2}\pi^{-1/2}\Gamma \big(\frac{r+1}{2}\big)$ denote the \textit{r}th absolute moment of an $\mathcal{N}(0,1)$ variable $\Phi$, where $\Gamma$ is the gamma function.\footnote{The gamma function is defined as $\Gamma(z) = \int_0^{\infty}t^{z-1} e^{-t}dt$.} We consider the family of realized power variations for arbitrary non-negative numbers $p,q\in\mathbb{R}$:
\begin{equation}
    B^n(p,q)_t = \sum_{i=1}^{N_t^n}\Delta(n,i)^{q+1-p/2}|\Delta_i^n Q|^p,
\end{equation}

\noindent where, in the regular sampling scheme, we have $B^n(p,q)=\Delta(n,i)^{q+1-p/2} B(p,\Delta_n)$, so all processes convey the same information when $q$ varies.

In the special case when $p=2$ and $q=0$, $B^n(2,0)$ is the approximate quadratic variation of the process $Q$ relative to the discretization scheme $S(n,i): i>0$, i.e.:
\begin{equation}
    B^n(2,0) = \sum_{i=1}^{N_t^n}|\Delta_i^n Q|^2.
\end{equation}
\noindent from which we have:
\begin{equation}
    B^n(2,0) \overset{u.c.p.}{\Rightarrow} m_2 C(2).
\end{equation}

\noindent where $C(2)=\int_0^t c_s ds$.
Analogously, when $p=4$ and $q=1$, $B^n(4,1)$ is an estimator for the quarticity:
\begin{equation}
    B^n(4,1) = \sum_{i=1}^{N_t^n}|\Delta_i^n Q|^4.
\end{equation}
A simplified version of Theorem 9.3 in \cite{aj14} assumes (H2) for $p=2$, and provides us with a Central Limit Theorem result for the asynchronous case. We then have that:
\begin{equation}
    \frac{\sqrt{m_4} (B^n(2,0)_T - m_2 C(2)_T)}{\sqrt{(m_4-m_2) B^n(4,1)_T}}  
\end{equation}

\noindent converges in law to $\mathcal{N}(0,1)$, conditional on the set $A \subset \{C_T>0\}$, with $P(A)>0$, for all $A\in \mathcal{F}$, with the Gaussian limit independent of the filtration $\mathcal{F}$. Given our null hypothesis, the conditioning on the set $A$ is immediately satisfied. 

Since we are performing a one-sided test, we choose the critical region to be:
\begin{equation}
    C_n = \bigg\{B^n(2,0)_T > m_2 \cdot c' T + z_\alpha \sqrt{1 - \frac{m_2}{m_4}}\cdot B^n(4,1)_T \bigg\}
\end{equation}

\noindent where the $B(p,q)^n$ are computed from $\Delta_i^n Q'$, instead of $\Delta_i^n Q$.

\vskip 12pt 

As we did for the regularly timed observations, we perform the test for absence of Brownian motion on asynchronous observations for the ten stocks (ABX, AC, BMO, BNS, ENB, MFC, RCI.B, RY, SU, and TD), for all traders listed in the Toronto Stock Exchange (TSX), for all trading days between January $2^{nd}$, 2010 to October $26^{th}$, 2020. We find that a Brownian component is relevant to the inventory process on the vast majority of days for active traders, as can be noted from Table \ref{tab:inventory_BM_async}. As before, we assume that on a given day the trader has traded on average at least once per minute to be included in the table. This is a very soft restriction for high-frequency trading.

\vskip 12pt

\begin{table}[H]
\centering
\begin{tabular}{lrrrrrrrrrr}
\hline
 &   ABX &    AC &    BMO &   BNS &    ENB &   MFC &  RCI.B &    RY &     SU &    TD \\
\hline
Trader $\#2$  &  98.5 &  99.5 &   98.6 &  98.6 &   98.6 &  97.4 &   99.2 &  97.9 &   97.6 &  97.6 \\
Trader $\#5$  &  91.5 &  97.4 &  100.0 &  91.8 &   91.5 &  85.6 &  100.0 &  81.9 &  100.0 &  89.9 \\
Trader $\#7$  &  98.4 &  99.5 &   98.2 &  97.5 &   97.6 &  96.3 &   98.6 &  96.8 &   97.3 &  94.3 \\
Trader $\#14$ &  98.6 &  99.3 &   98.9 &  99.2 &  100.0 &  97.9 &   99.3 &  99.1 &   98.3 &  98.4 \\
Trader $\#15$ &  98.4 &  97.7 &   98.4 &  98.3 &   97.6 &  97.6 &   98.2 &  98.2 &   97.3 &  98.4 \\
Trader $\#39$ &  96.9 &  98.1 &   95.6 &  96.3 &   98.1 &  97.1 &   97.4 &  97.0 &   97.6 &  97.1 \\
Trader $\#53$ &  95.5 &  94.8 &   96.2 &  95.4 &   96.9 &  95.4 &   96.4 &  95.7 &   98.1 &  96.6 \\
Trader $\#65$ &  98.3 &  99.2 &   98.2 &  98.9 &   98.4 &  98.1 &   98.9 &  98.7 &   98.1 &  98.4 \\
\hline
\end{tabular}
\caption{Percentage of days for which there is Brownian component in asynchronous observations for the inventory data, active traders.}
\label{tab:inventory_BM_async}
\end{table}

Table \ref{tab:wealth_BM_asynch} presents the same conclusion for the case of asynchronous observations of the wealth process. Namely, that we must model the process by including the diffusion term, instead of considering the process to be strictly time dependent.

\vskip 12pt

\begin{table}[H]
\centering
\begin{tabular}{lrrrrrrrrrr}
\hline
 &   ABX &    AC &    BMO &   BNS &    ENB &   MFC &  RCI.B &    RY &     SU &    TD \\
\hline
Trader $\#2$  &   97.0 &   98.0 &   94.9 &   95.5 &  94.1 &   95.3 &   96.2 &   95.5 &   95.0 &   94.5 \\
Trader $\#5$  &  100.0 &  100.0 &  100.0 &  100.0 &  99.9 &  100.0 &  100.0 &  100.0 &  100.0 &  100.0 \\
Trader $\#7$  &   97.3 &   99.0 &   95.8 &   95.0 &  94.3 &   95.2 &   95.2 &   94.8 &   95.5 &   92.4 \\
Trader $\#14$ &   97.4 &   96.7 &   96.6 &   95.3 &  93.0 &   84.7 &   95.8 &   97.4 &   92.1 &   94.1 \\
Trader $\#15$ &   96.6 &   99.1 &   96.4 &   95.6 &  96.1 &   96.8 &   95.5 &   97.0 &   95.4 &   94.7 \\
Trader $\#39$ &   94.6 &   96.7 &   94.0 &   92.0 &  93.4 &   94.4 &   95.2 &   93.7 &   93.1 &   93.7 \\
Trader $\#53$ &   93.5 &   97.1 &   92.8 &   90.6 &  93.0 &   92.9 &   95.1 &   91.5 &   90.4 &   91.0 \\
Trader $\#65$ &   97.0 &   99.5 &   95.7 &   96.1 &  96.7 &   96.6 &   96.2 &   94.8 &   94.0 &   96.7 \\
\hline
\end{tabular}
\caption{Percentage of days for which there is Brownian component in asynchronous observations for the inventory data, active traders.}
\label{tab:wealth_BM_asynch}
\end{table}

Hence, we conclude that \textit{we should model the inventory $Q$ and wealth $X$ processes using Brownian motion terms for the irregularly observed processes as well.}

\section{Optimal Execution with Unbounded Variation Inventory}
\label{sec:3}

The previous section provided convincing arguments for the inventories and wealth processes of traders to be It\^o processes with non-trivial Brownian components. The goal of this section is to investigate how this should affect the solutions of optimal execution problems. Instead of working with the time honored approach within which inventory processes (and hence wealth processes) are differentiable in time, we here assume that the inventory of the trader in charge of the execution has a Brownian motion component.  
We use an extension of a solvable model introduced in \cite{cj16} (see also \cite{LL1}) as a test bed for our new paradigm.

\subsection{Theoretical Analysis of the Model}
\label{sec:3.1}
Using the same notation as \cite{LL1}, our goal is to compute the value function $V(t_0,x,s,q)$ of the trading agent as well as the control permitting to actually realize this optimal value.
Here $t_0$ is the time at which this value is computed and  at that time, $x$ is the wealth of the agent, $q$ is the inventory (number of shares) they hold, and $s$ is the mid-price of the traded security. The wealth is understood as the amount of cash held by the agent plus the value of the share inventory marked at the mid-price. For the sake of definiteness, we follow \cite{cj16} and define the value function $V$ as the result of the optimization:
\begin{equation}
\label{fo:value}
V(t_0,x,s,q) = \sup_{(\nu_t)_t} \EE \Big[ X_T + Q_T S_T-A |Q_T-q_T|^2 -\phi \int_{t_0}^T |Q_t|^2 dt \;\; \Bigl|\;\; X_{t_0}=x,S_{t_0}=s,Q_{t_0}=q\Big]
\end{equation}
where $\nu_t$ represents the speed of trading for the agent at time $t$. By convention, $\nu_t>0$ when the agent is buying and $\nu_t<0$ when they are selling. This is the control of the trader. The processes appearing in the above definition are assumed to have the following It\^o dynamics:
\begin{equation}
\label{fo:oe}
\begin{cases}
dS_t = \alpha \nu_t dt + \sigma dW_t,\quad S_t>0 \;\;t\ge 0\;\; S_{t_0}=s\\
dQ_t = \nu_t dt + \tilde{\sigma} d\widetilde{W}_t,\;\;Q_{t_0}=q,\\
dX_t = -\nu_t(S_t+\kappa \nu_t)dt-\tilde\sigma(S_t+\kappa\nu_t) d\tilde W_t,\;\;X_{t_0}=x,\\
\end{cases}
\end{equation}
The meanings and the significance of the constants $\alpha$, $\kappa$, $\sigma$ and $\tilde\sigma$ can be inferred from the above equations giving the dynamics of the mid-price $S_t$, the inventory $Q_t$ and the cash $X_t$ of the agent. The positive constants $\alpha$ and $\kappa$ capture the effects of the permanent and temporary price impacts which we assume to be linear in the speed of trading for the sake of simplicity. As for $\sigma$ and $\tilde\sigma$, they obviously stand for the volatilities of the mid-price and the inventory, respectively. The main difference with the model used in \cite{cj16} and \cite{LL1} is the fact that the dynamics of the inventory $Q_t$ and the wealth $X_t$ contain Brownian martingale terms. First, we added the term $\tilde\sigma d\widetilde{W}_t$ to the dynamics of the inventory. Second, for consistency reasons, we keep the relationship $dX_t=-(S_t+\kappa\nu_t)dQ_t$, and using the new form of $dQ_t$ in the second equation of \eqref{fo:oe} we end up with the dynamics of the wealth in the third equation of \eqref{fo:oe}. Note also that the above form of the optimal execution problem is a relaxed form of an optimal liquidation problem. Indeed, in practical execution problems, the agent aims at a specific inventory at the terminal time $T$. Hence, formula \eqref{fo:oe} should include a \emph{hard constraint} of the form $Q_T=q_T$ where the real number $q_T$ is the desired inventory at time $T$. Instead, the model we are using includes a penalty $A|Q_T-q_T|^2$ for a positive constant $A$ to favor trading strategies with final inventory as close to the target as possible.
For the sake of definiteness, we here chose to solve an optimal liquidation (resp. acquisition) problem, in which case $q_T=0$ and the initial inventory $Q_0=q$ is the quantity of assets to be disposed of (resp. acquired) by time $T$.

\begin{remark}
\label{re:hypotheses}
According to the discussion  of adverse selection given in \cite{CarmonaWebster}, the correlation between the Brownian motions driving the first two equations in \eqref{fo:oe} should be negative when trading with market orders and positive when trading with limit orders. Trying not to constrain the trader to a specific type of order, we assume that these two Brownian motions are independent. Note also that since we do not posit an order book model and we do not distinguish between the cash and share contributions to the wealth, the dynamics of the wealth in \eqref{fo:oe} is slightly different from the the self-financing condition in \cite{CarmonaWebster}. In the spirit of the \emph{transaction costs literature}, we choose to follow \cite{cj16} and only model the transaction costs for crossing the spread. Moreover, like in \cite{cj16} and \cite{LL1}, we assume that this spread is proportional to the rate of trading.
\end{remark}

\vskip 6pt
If it were not for the lack of concavity of the terminal reward and the fact that the dynamics are quadratic in the control, the above optimal control problem could be treated as a linear quadratic model and solved explicitly, despite the fact that the volatility depends upon the control. Like in \cite{cj16} we solve this control problem by guessing the form of the solution. However, instead of using an analytic approach based on the solution of the HJB equation and using a verification argument to conclude, we choose an alternative approach based on the stochastic maximum principle, in no small part because of our desire to identify the optimal control.
The Hamiltonian of the problem is given by:
$$
H(t,\bx,\by,\bz,\nu)=\alpha\nu y^s+\nu y^q-\nu(s+\kappa\nu)y^x+\sigma z^s+\tilde \sigma z^q - \tilde\sigma(s+\kappa\nu)\tilde z^x+\phi q^2
$$
if we denote the state by $\bx=(s,q,x)$ and the adjoint variables (also called co-variates) by $\by=(y^s,y^q,y^x)$ and $\bz=(z^s,\tilde z^s,z^q,\tilde z^q,z^x,\tilde z^x)$.
For each admissible control process $(\nu_t)_{0\le t\le T}$ the stochastic process 
$$
(\bY_t,\bZ_t)=\bigl((Y_t^s,Y_t^q,Y_t^x),(Z_t^s,\tilde Z_t^s, Z_t^q,\tilde Z_t^q, Z_t^x,\tilde Z_t^x)\bigr)
$$ 
is called the associate adjoint process if it satisfies the following Backward Stochastic Differential equation:
\begin{equation}
    \label{fo:adjoint_equation}
    \begin{cases}
dY^s_t=(\nu_t Y^x_t +\tilde\sigma \tilde Z^x_t)dt +Z^s_t dW_t + \tilde Z^s_t d\tilde W_t,\qquad Y^s_T=-Q_T\\
dY^q_t=-2\phi Q_t dt +Z^q_t dW_t + \tilde Z^q_t d\tilde W_t,\qquad Y^q_T=2AQ_T-S_T\\
dY^x_t=Z^x_t dW_t + \tilde Z^x_t d\tilde W_t, \qquad Y^x_T=-1\\
    \end{cases}
\end{equation}
The third equation of \eqref{fo:adjoint_equation} is easy to solve. Its solution is given by $Z^x_t=\tilde Z^x_t= 0$ so that $Y^x_t= -1$. Knowing that $\tilde Z^x_t= 0$, the first equation can also be taken care of by choosing $Z^s_t= 0$ and $\tilde Z^s_t= -\tilde\sigma$ and $Y^s_t\equiv - Q_t$ which turns this first equation into the equation satisfied by $Q_t$ in \eqref{fo:oe}.

In order to solve the second equation of \eqref{fo:adjoint_equation}, we make the ansatz
\begin{equation}
    \label{fo:ansatz1}
Y^q_t=-S_t+v(t,Q_t),
\end{equation}
in which case the unknown function $(t,q)\mapsto v(t,q)$ should also satisfy the terminal condition $v(T,q)=2Aq$. Assuming that $v$ is differentiable enough, we compute the stochastic differential of $Y^q_t$ using It\^o formula and \eqref{fo:oe}:
\begin{equation}
    \begin{split}
dY^q_t
&=-dS_t+\partial_t v(t,Q_t)dt+\partial_q v(t,Q_t)dQ_t+\frac12 \partial^2_{qq}v(t,Q_t)d[Q,Q]_t\\
&=-\alpha \nu_t dt - \sigma dW_t
+\partial_t v(t,Q_t)dt+\partial_q v(t,Q_t)[\nu_t dt + \tilde{\sigma} d\widetilde{W}_t]+\frac{\tilde\sigma^2}{2} \partial^2_{qq}v(t,Q_t)dt\\
&=[-\alpha \nu_t+\partial_t v(t,Q_t)+\nu_t\partial_q v(t,Q_t)+\frac{\tilde\sigma^2}{2} \partial^2_{qq}v(t,Q_t)]dt
- \sigma dW_t +\tilde{\sigma}\partial_q v(t,Q_t) d\widetilde{W}_t
    \end{split}
\end{equation}
so that \eqref{fo:ansatz1} will give a solution to the second equation of \eqref{fo:adjoint_equation} if we choose $Z^q_t=-\sigma$, $\tilde Z^q_t=\tilde\sigma\partial_qv(t,Q_t)$ and the function $v$ such that 
\begin{equation}
\label{fo:v1}
-\alpha \nu_t+\partial_t v(t,Q_t)+\nu_t\partial_q v(t,Q_t)+\frac{\tilde\sigma^2}{2} \partial^2_{qq}v(t,Q_t)=-2\phi Q_t.
\end{equation}
Finding such a function $v$ would give us an explicit construction of the adjoint process associated to a general control $\nu_t$. However, we are mostly interested in the optimal control which we shall denote by $\hat\nu_t$.
With this in mind, we use the fact that the necessary part of the stochastic maximum principle states that if an admissible control process $(\nu_t)_{0\le t\le T}$ is optimal, and if $\bigl(\bY,\bZ\bigr)$ is the associated adjoint process, then the Hamiltonian is minimized along the trajectories. So we search for a minimizer of the Hamiltonian with respect to the variable $\nu$. The first order condition $\partial_\nu H=0$ reads
$$
\alpha y^s+y^q-2\kappa y^x\nu -sy^x-\kappa\tilde\sigma\tilde z^x=0
$$
which is solved by
\begin{equation}
    \label{fo:nu_hat}
\hat\nu=\frac{-sy^x+\alpha y^s+y^q-\kappa\tilde\sigma\tilde z^x}{2\kappa y^x},
\end{equation}
the constraint $y^x<0$ being added to guarantee that the critical point $\hat\nu$ is in fact a minimum. This implies that the optimal control should be given by:
\begin{equation}
    \label{fo:nu_hat_t}
\begin{split}
\hat\nu_t&=\frac{-S_tY^x_t+\alpha Y_t^s+Y_t^q}{2\kappa Y_t^x}\\
&=\frac{S_t-\alpha Q_t -S_t+v(t,Q_t)}{-2\kappa}\\
&=\frac{\alpha}{2\kappa} Q_t -\frac{1}{2\kappa}v(t,Q_t).
\end{split}
\end{equation}
Recall that $\tilde Z^x_t=0$ and that $Y^x_t=-1$ satisfies the negativity constraint.
Injecting this expression into \eqref{fo:v1}, we see that we can determine the process $Y^q_t$ at the optimum if we can find a function $v$ satisfying the partial differential equation:
$$
-\alpha \Bigl( \frac{\alpha}{2\kappa} q -\frac{1}{2\kappa}v(t,q)\Bigr)
+\partial_t v(t,q)
+\Bigl(\frac{\alpha}{2\kappa} q -\frac{1}{2\kappa}v(t,q)\Bigr)\partial_q v(t,q)
+\frac{\tilde\sigma^2}{2} \partial^2_{qq}v(t,q)
=-2\phi q,
$$    
or in other words
\begin{equation}
    \label{fo:pde}
\partial_t v(t,q) 
+\frac{\tilde\sigma^2}{2} \partial^2_{qq}v(t,q)
+\Bigl(\frac{\alpha}{2\kappa} q -\frac{1}{2\kappa}v(t,q)\Bigr)\partial_q v(t,q)
+\frac{\alpha}{2\kappa}v(t,q) +\Bigl(2\phi-\frac{\alpha^2}{2\kappa}\Bigr) q=0,
\end{equation}
with the terminal condition $v(T,q)=2Aq$.
Given the special form of this partial differential equation, we search for a solution in the form
\begin{equation}
    \label{fo:ansatz2}
    v(t,q)=\eta_t q+\chi_t,
\end{equation}
for some deterministic functions $t\mapsto\eta_t$ and $t\mapsto\chi_t$.
Injecting \eqref{fo:ansatz2} into \eqref{fo:pde} we find that the function $v$ defined by \eqref{fo:ansatz2} solves \eqref{fo:pde} if and only if the functions $\eta_t$ and $\chi_t$ solve the system:
\begin{equation}
    \label{fo:odes}
\begin{cases}
\dot\eta_t&=-\frac{\alpha}{\kappa}\eta_t+\frac1{2\kappa}\eta_t^2+\Bigl(\frac{\alpha^2}{2\kappa}-2\phi\Bigr),\qquad \eta_T=2A,\\
\dot\chi_t&=\frac{\eta_t-\alpha}{2\kappa}\chi_t,\qquad \chi_T=0.
\end{cases}
\end{equation}
The first equation is a Riccati equation. Notice that if $\alpha^2<4\kappa\phi$, then this equation has a unique solution given by:

\begin{equation}
    \label{fo:eta_t}
\eta_t=\frac{\Bigl(\frac{\alpha^2}{2\kappa}-2\phi\Bigr)\Bigl(e^{2\sqrt{\phi/\kappa}(T-t)}-1\Bigr) -2A\Bigl(\Bigl(\frac{\alpha}{2\kappa}+\sqrt{\frac{\phi}{\kappa}}\Bigr)
e^{2\sqrt{\phi/\kappa}(T-t)}-\Bigl(\frac{\alpha}{2\kappa}-\sqrt{\frac{\phi}{\kappa}}\Bigr)\Bigr) }
{\Bigl(\frac{\alpha}{2\kappa}-\sqrt{\frac{\phi}{\kappa}}\Bigr)e^{2\sqrt{\phi/\kappa}(T-t)}-\Bigl(\frac{\alpha}{2\kappa}+\sqrt{\frac{\phi}{\kappa}}\Bigr)
-\frac{A}{\kappa}\Bigl(e^{2\sqrt{\phi/\kappa}(T-t)}-1\Bigr)}.
\end{equation}
Once the function $\eta_t$ is computed, we can inject its value into the second equation in \eqref{fo:odes} to see that this equation appears as a first order linear equation with $0$ terminal condition, so its solution is identically $0$, i.e. $\chi_t=0$.

\subsection{Practical Implementation}
\label{sec:3.2}
For the purpose of illustration we choose a stock which is actively traded on the Toronto Stock Exchange (TSX). We chose Royal Bank of Canada\footnote{Royal Bank of Canada is a Canadian multinational financial services company and the largest bank in Canada by market capitalization.} with symbol RY. One of the characteristics of the data is that as in most data in TAQ format, a price and a volume are reported for each transaction. but contrary to most exchanges, the identities of the buyer and the seller are also provided. So we can choose any of the brokers registered on TSX and follow their trades on RY during the day. We choose the specific brokerage firm Citadel Securities Canada ULC for two specific reasons: 1) they were actively trading RY on the days 03/25/2020 and 08/27/2020 which we chose for the purpose of illustration, for example having executed 2357 trades on 03/25/2020, and 2) it was clear on both days, what the target for their inventory at the end of the trading day was. On 03/25/2020 their inventory was relatively small at the end of the day, with only $144$ shares remaining in the inventory was. For that reason, we shall assume $q=q_T=0$. On 08/27/2020, they tried to acquire a large number of shares. So, in order to avoid having to recalibrate the constant parameters involved in the cost function, we keep the terminal target of $q=q_T=0$, and we adjust their initial inventory to be the negative of their actual inventory at the end of the day.

\vskip 2pt
We denote by $\tau_1<\tau_2<\tau_3<\ldots<\tau_N$ the times at which Citadel was involved in a trade, and for each integer $i$, $1\le i\le N$ we denote by \begin{itemize}
    \item $s_i=S_{\tau_i}$ the price at which the transaction involving Citadel took place; 
    \item $q_i$ the inventory of Citadel right after the transaction at time $\tau_i$. So $q_i$ is the algebraic sum of numbers of shares bought or sold at times $\tau_1$, $\tau_2$, $\cdots$, $\tau_i$. In other words, $q_i-q_{i-1}$ is equal to the quantity bought or sold at time $\tau_i$;
    \item $x_i$ the wealth of Citadel right after the transaction at time $\tau_i$. So $x_i$ is the algebraic sum of the payments in the transactions at times $\tau_1$, $\tau_2$, $\cdots$, $\tau_i$. In other words, $x_i-x_{i-1}$ is equal to the quantity bought or sold at time $\tau_i$ times the price, i.e. $(q_i-q_{i-1})s_i$, and $x_0=0$.
\end{itemize}
Next, we use \cite{LL1} to estimate the remaining parameters of the model \eqref{fo:oe}, namely $\hat A$, $\hat \phi$, $\hat\alpha$, $\hat\sigma$, $\hat\kappa$. Details and numerical estimates are given in the Appendix. Note that $\tilde\sigma$ is the only new parameter to estimate. We choose a naïve empirical estimate of the standard deviation of the series $(q_i-q_{i-1})_{1\le i\le N}$ defined above.

\vskip 6pt
Our goal in this section is to compare the actual terminal inventory and terminal wealth of Citadel to what could have been obtained by using the optimal execution policy identified in the previous section. Because of the presence of the Brownian motions $(W_t)_{t\ge 0}$ and 
$(\tilde W_t)_{t\ge 0}$ in the equations defining the model, we can only compare the inventory and the wealth of Citadel to their expected values should they had followed the optimal trading rate identified above in equation \eqref{fo:nu_hat_t}. While this is a desirable goal, the comparison we actually perform is marred by three approximations: 1) while the optimal rate of trading is for continuous trading, we implement the optimal trading strategy at the times at which Citadel traded on that day; 2) we use a Monte Carlo approximation of the expectations; 3) by trading a different amount than Citadel, the trades of our simulated scenarios should not have the same price impact as the actual price impact incurred by Citadel, and we make an educated guess to account for this difference.

\vskip 6pt
We propose two possible approaches based on the three points above. In each case we use a simple Euler discretization of the second and third equations in \eqref{fo:oe} at the times $\tau_i$ to update the values of the inventory and the wealth. In all cases we shall use $N_{sim}$ Monte Carlo simulations, say $N_{sim}=10,000$, of the increments of the Brownian motions.

\subsubsection{Approach \#1}

In this approach, we do not use the time evolution of the price $S_t$ as given by the first equation in \eqref{fo:oe}. We only use the Brownian motion $(\tilde W_t)_{t\ge 0}$ appearing in the second  equation of \eqref{fo:oe}, and in the update of the wealth from time $\tau_{i-1}$ to $\tau_i$, we use the actual price $S_{\tau_{i-1}}$.
\begin{itemize}
    \item For each  $j=1,2,\cdots,N_{sim}$, let $(\tilde\epsilon_i^j)_{1\le i\le N}$ be a sequence of $N$ independent $N(0,1)$ random variables. We use $\sqrt{\tau_i-\tau_{i-1}}\tilde\epsilon^j_i$ for the increment $d\tilde W_t$ of the $j$-sample of $(\tilde W_t)_{t\ge0}$ between the times $\tau_{i-1}$ and $\tau_i$.
    \item For each  $j=1,2,\cdots,N_{sim}$, we construct the sequences $(q_i^j)_{1\le i\le N}$ and $(x_i^j)_{1\le i\le N}$ giving respectively the values of the $j$-th sample of the inventory process $(Q_t)_{t\ge 0}$ and the wealth process $(X_t)_{t\ge 0}$ at the times $\tau_i$ recursively following the steps:
    \begin{itemize}
        \item given $q^j_{i-1}$, we define $q^j_i=q^j_{i-1}+\nu^j_{i-1}(\tau_i-\tau_{i-1})+\hat{\tilde\sigma}\sqrt{\tau_i-\tau_{i-1}}\tilde\epsilon^j_i$ where $\nu^j_{i-1}$ is the value of $\hat\nu_{\tau_{i-1}}$ computed as
        $$
        \nu^j_{i-1}=\frac{\hat\alpha}{2\hat\kappa}q^j_{i-1}-\frac{1}{2\hat\kappa}v(\tau_{i-1},q^j_{i-1})
        $$
       with $v(\tau_{i-1},q^j_{i-1})=\eta_{\tau_{i-1}}q^j_{i-1}$ and $\eta_t$ is given by formula \eqref{fo:eta_t}. 
        \item given $x^j_{i-1}$, we define $x^j_i=x^j_{i-1}-\nu^j_{i-1}s_{i-1}(\tau_i-\tau_{i-1})-\hat{\tilde\sigma}s_{i-1}\sqrt{\tau_i-\tau_{i-1}}\tilde\epsilon^j_i$;\\
        Notice that since the price observed is the trade price, not the mid price, we do not add the $\kappa \nu$ component which corresponds to crossing the spread.
    \end{itemize}
    \item for each $i=1,2,\cdots,N$, we can draw on the same plot the actual inventory $q_i$ of Citadel at time $\tau_i$ and on the same vertical line over $t=\tau_i$, the $5\%$-tile and the $95\%$-tile of the $(q^j_i)_{1\le j\le N_{sim}}$. We can draw the similar plot comparing the actual wealth $x_i$ of Citadel at time $\tau_i$ to the band of the optimal wealth scenarios.
    \item Still, the easiest statistics to compute are the average terminal wealth 
    $$
    \bar x_N=\frac{1}{N_{sim}}\sum_{j=1}^{N_sim}x^j_N
    $$ 
and the percentage of Monte Carlo scenarios of the wealth which end up at time $T$ with a larger value than the actual wealth  accumulated by Citadel for trading on RY on that day.
\end{itemize}

Beyond the Monte Carlo approximations inherent in the Euler simulations of 
the equations of motion \eqref{fo:oe}, the major approximation in the above comparisons is the fact that the stock price at time $\tau_i$ is the price $s_i$ experienced by Citadel after trading the quantity $q_i-q_{i-1}$, while the stock price should have been affected by the price impact produced by the volume $\nu^j_i(\tau_i-\tau_{i-1})$. The next approach is an attempt to reconcile the procedure outlined in the above bullet points with the equation in \eqref{fo:oe} giving the evolution of the stock price from the price impact due to the trading rate and the random shocks coming from the Brownian motion $(W_t)_{t\ge 0}$.

\subsubsection{Approach \#2}

\begin{itemize}
    \item For each  $j=1,2,\cdots,N_{sim}$, we now use two independent sequences  $(\epsilon_i^j)_{1\le i\le N}$ and $(\tilde\epsilon_i^j)_{1\le i\le N}$ each consisting of $N$ independent $N(0,1)$ random variables. We will use $\sqrt{\tau_i-\tau_{i-1}}\epsilon^j_i$ (resp. $\sqrt{\tau_i-\tau_{i-1}}\tilde\epsilon^j_i$) for the increment $dW_t$ (resp. $d\tilde W_t$) of the $j$-sample of $(W_t)_{t\ge0}$ (resp. $(\tilde W_t)_{t\ge0}$) between the times $\tau_{i-1}$ and $\tau_i$.
    \item For each  $j=1,2,\cdots,N_{sim}$, we construct the sequences $(s_i^j)_{1\le i\le N}$, $(q_i^j)_{1\le i\le N}$ and $(x_i^j)_{1\le i\le N}$ giving respectively the values of the $j$-th sample of the price process $(S_t)_{t\ge 0}$, the inventory process $(Q_t)_{t\ge 0}$ and the wealth process $(X_t)_{t\ge 0}$ at the times $\tau_i$ recursively following the steps:
    \begin{itemize}
        \item given $q^j_{i-1}$, we define $q^j_i=q^j_{i-1}+\nu^j_{i-1}(\tau_i-\tau_{i-1})+\hat{\tilde\sigma}\sqrt{\tau_i-\tau_{i-1}}\tilde\epsilon^j_i$ where $\nu^j_{i-1}$ is the value of $\hat\nu_{\tau_{i-1}}$ computed as
        $$
        \nu^j_{i-1}=\frac{\hat\alpha}{2\hat\kappa}q^j_{i-1}-\frac{1}{2\hat\kappa}v(\tau_{i-1},q^j_{i-1})
        $$
       with $v(\tau_{i-1},q^j_{i-1})=\eta_{\tau_{i-1}}q^j_{i-1}$ and $\eta_t$ is given by formula \eqref{fo:eta_t}. 
       \item given $s^j_{i-1}$, we define $s^j_i=s^j_{i-1}+\hat\alpha\nu^j_{i-1}(\tau_i-\tau_{i-1})+\hat{\sigma}\sqrt{\tau_i-\tau_{i-1}}\epsilon^j_i$ where $\nu^j_{i-1}$ is defined in the above bullet point.
       \item given $x^j_{i-1}$, we define $x^j_i=x^j_{i-1}-\nu^j_{i-1}(s^j_{i-1}+\hat\kappa\nu^j_{i-1})(\tau_i-\tau_{i-1})-\hat{\tilde\sigma}(s^j_{i-1}+\hat\kappa\nu^j_{i-1})\sqrt{\tau_i-\tau_{i-1}}\tilde\epsilon^j_i$;
    \end{itemize}
    \item we can then proceed as before to plot both for the inventory and the wealth, the actual and simulated scenarios, and we can also compute the same summary statistics as in Approach \#1.
\end{itemize}

\begin{remark}
\label{re:cj1}
Notice that setting $\tilde\sigma=0$ corresponds to the original model in \cite{cj16}. In that case, all the $N_{sim}$ Monte Carlo simulations of the optimal inventory and the optimal wealth are identical in Approach \#1, and in some sense, they give the optimal inventory and the optimal wealth of the trader, had they traded at the same times $\tau_i$ as Citadel, had they bought and sold at these times the quantities prescribed by the solution of the optimal execution problem given in \cite{cj16}, and assuming that they could have traded these quantities at the prices $s_i$, in other words assuming that the price impact due to their trades would be no different than the price impact affecting the trade of the quantity $q_i-q_{i-1}$. So if we are willing to ignore this last detail, we have a way to compare the actual behavior of a trader on the floor, to the optimal behavior prescribed by the model to reach the same terminal inventory. 
\end{remark}

\subsection{Numerical Examples}
\label{sec:3.3}

For the numerical illustrations provided below, we generated $N_{sim}=10,000$ Monte Carlo simulations of the \emph{optimal} inventories and wealths in each of the two approaches detailed above. At each time $\tau_i$ of the trades take place we computed the $5\%$-tile and the $90\%$-tile of these scenarios and we plotted in light blue the band limited by these two percentiles. Also for the purpose of illustration, we plotted the actual trajectories of the first $5$ scenarios. 

The illustrations were produced using the numeric value $\sigma_S$ estimated from the past ten days of asynchronous intraday trading for the volatility of the price. And the value $\sigma_Q=\tilde \sigma$ from the same day volatility of the trader's inventory. On 08/27/2020, we have $\sigma_S=0.21$ and $\sigma_Q=25707.43$, while on 03/25/2020, $\sigma_S=1.49$ and $\sigma_Q=1449.32$, respectively.

\begin{figure}[H]
  \textbf{Inventory Evolution in Both Approaches}\par\medskip
\centerline{
\includegraphics[width=6.5cm,height=6.5cm]{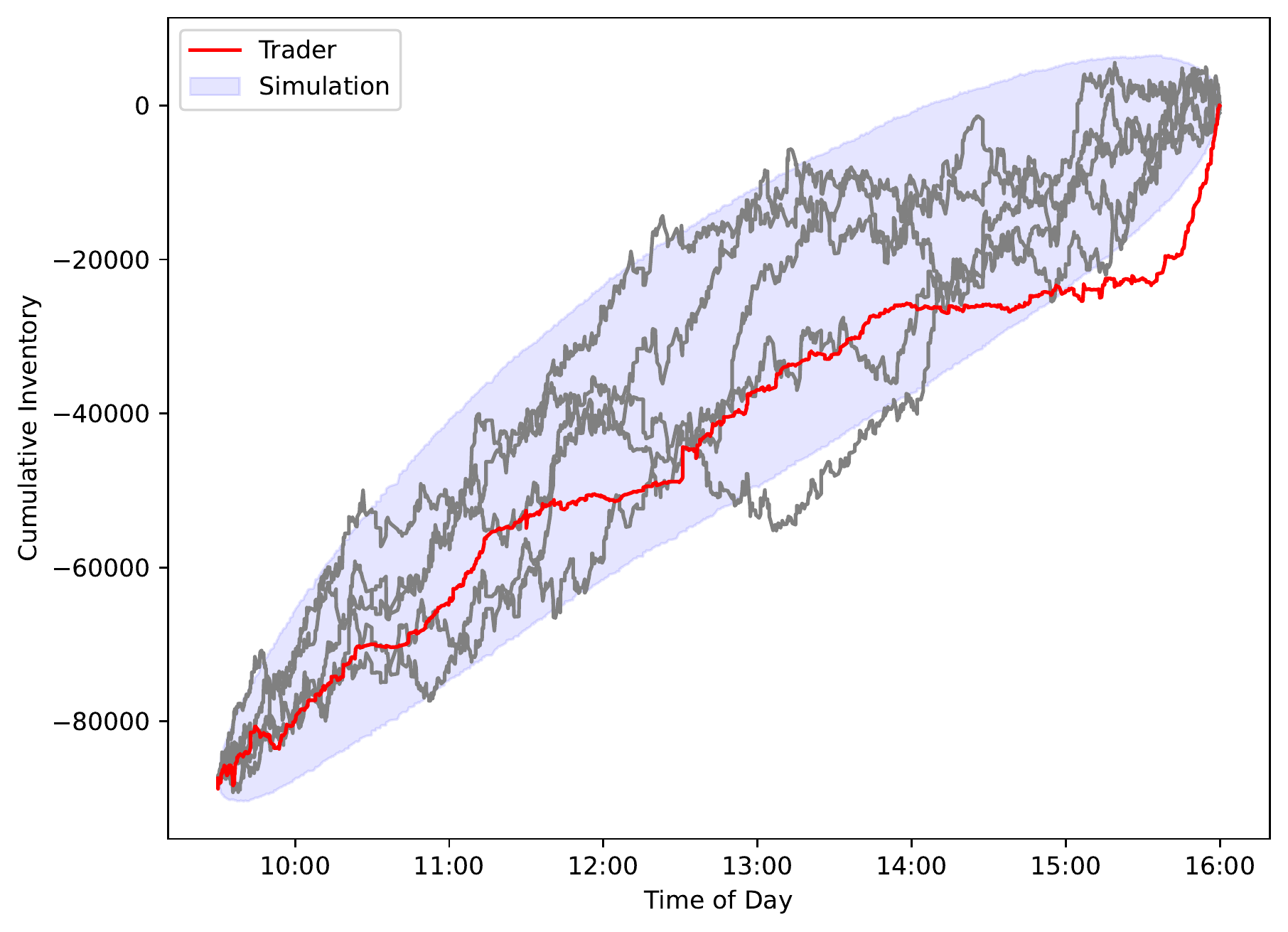}
\includegraphics[width=6.5cm,height=6.5cm]{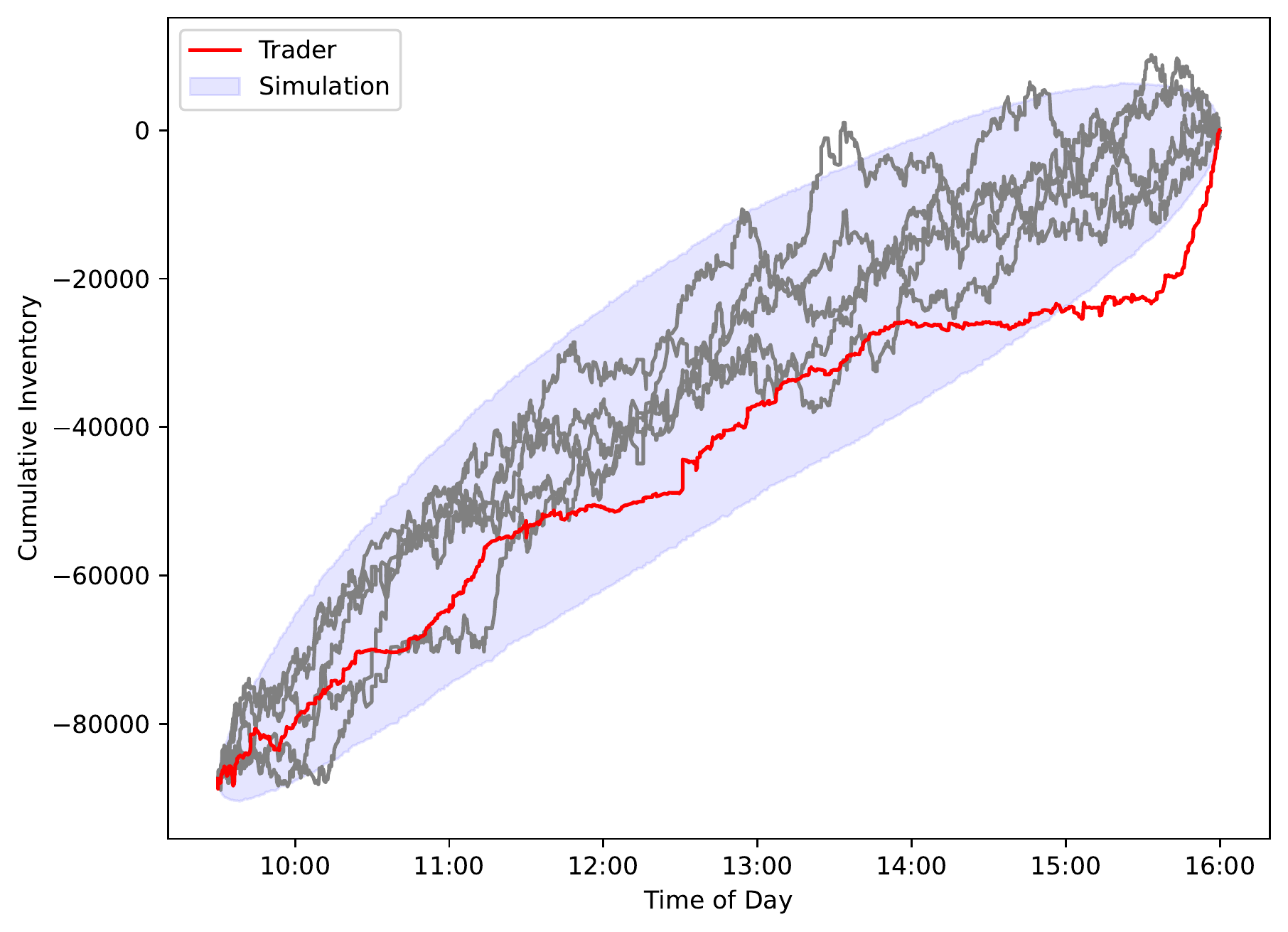}
}
\label{fi:inventory082720}
\caption{Plot of the actual RY inventory (red curve) of Citadel on 08/27/2020 together with $5$ sample Monte Carlo scenarios of what the inventory would have been had the optimal rate of trading been used. The light blue band is bounded by the $5\%$-tile and the $90\%$-tile of $N_{sim}=10,000$ scenarios. The left pane gives the result of Approach \#1 and the plot in the right pane was produced following Approach \#2.} 
\end{figure}

\begin{figure}[H]
  \textbf{Wealth Evolution in Both Approaches}\par\medskip
\centerline{
\includegraphics[width=6.5cm,height=6.5cm]{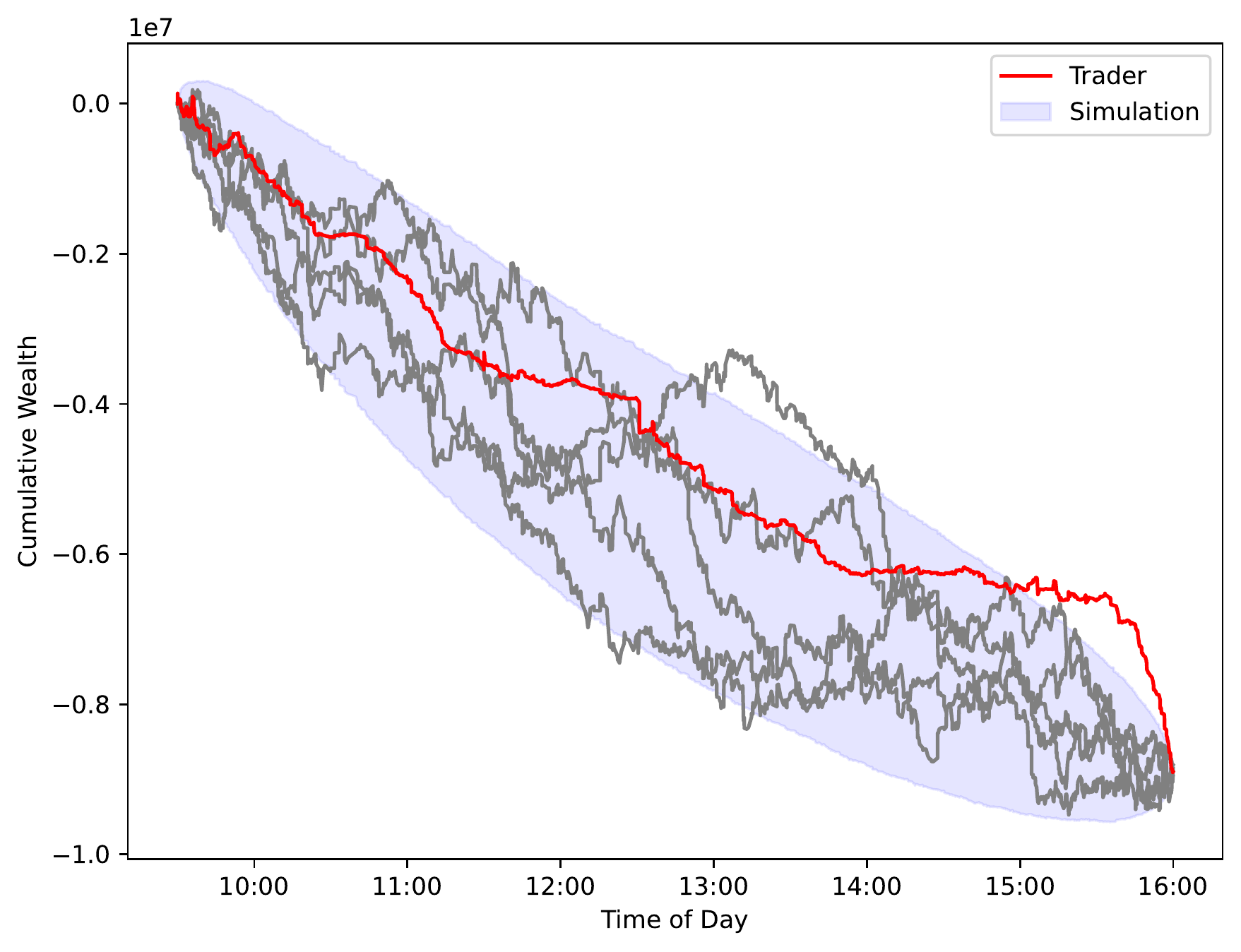}
\includegraphics[width=6.5cm,height=6.5cm]{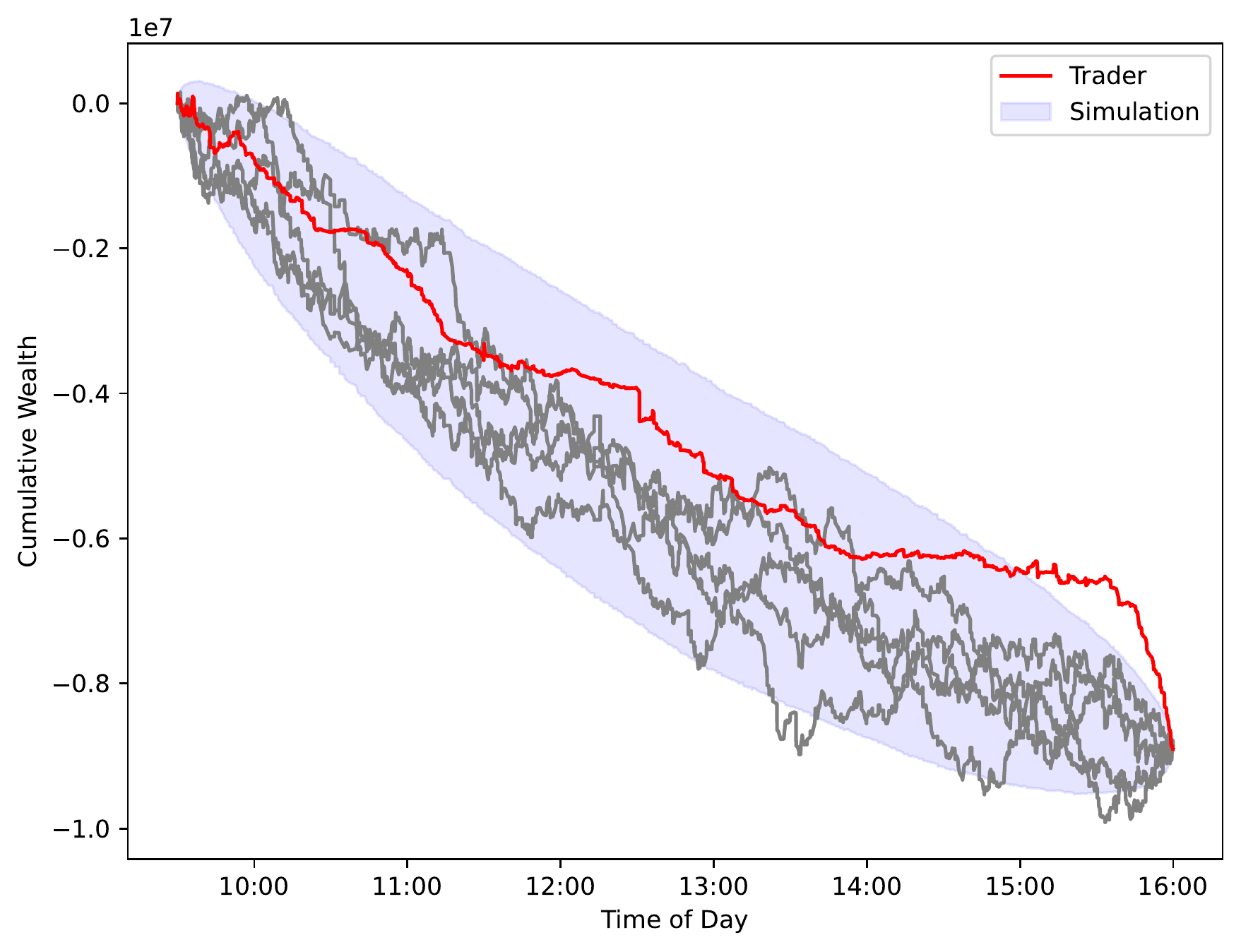}
}
\caption{Plot of the actual wealth (red curve) accumulated by Citadel by trading RY on 08/27/2020 together with $5$ sample Monte Carlo scenarios of what this wealth would have been had the optimal rate of trading been used. The light blue band is bounded by the $5\%$-tile and the $90\%$-tile of $N_{sim}=10,000$ scenarios. The left pane gives the result of Approach \#1 and the plot in the right pane was produced following Approach \#2.} 
\label{fi:wealth082720}
\end{figure}

\begin{figure}[H]
\textbf{Terminal Wealth Comparison, Kernel Density}\par\medskip
\centering
\begin{minipage}{.45\linewidth}
  \includegraphics[width=\linewidth]{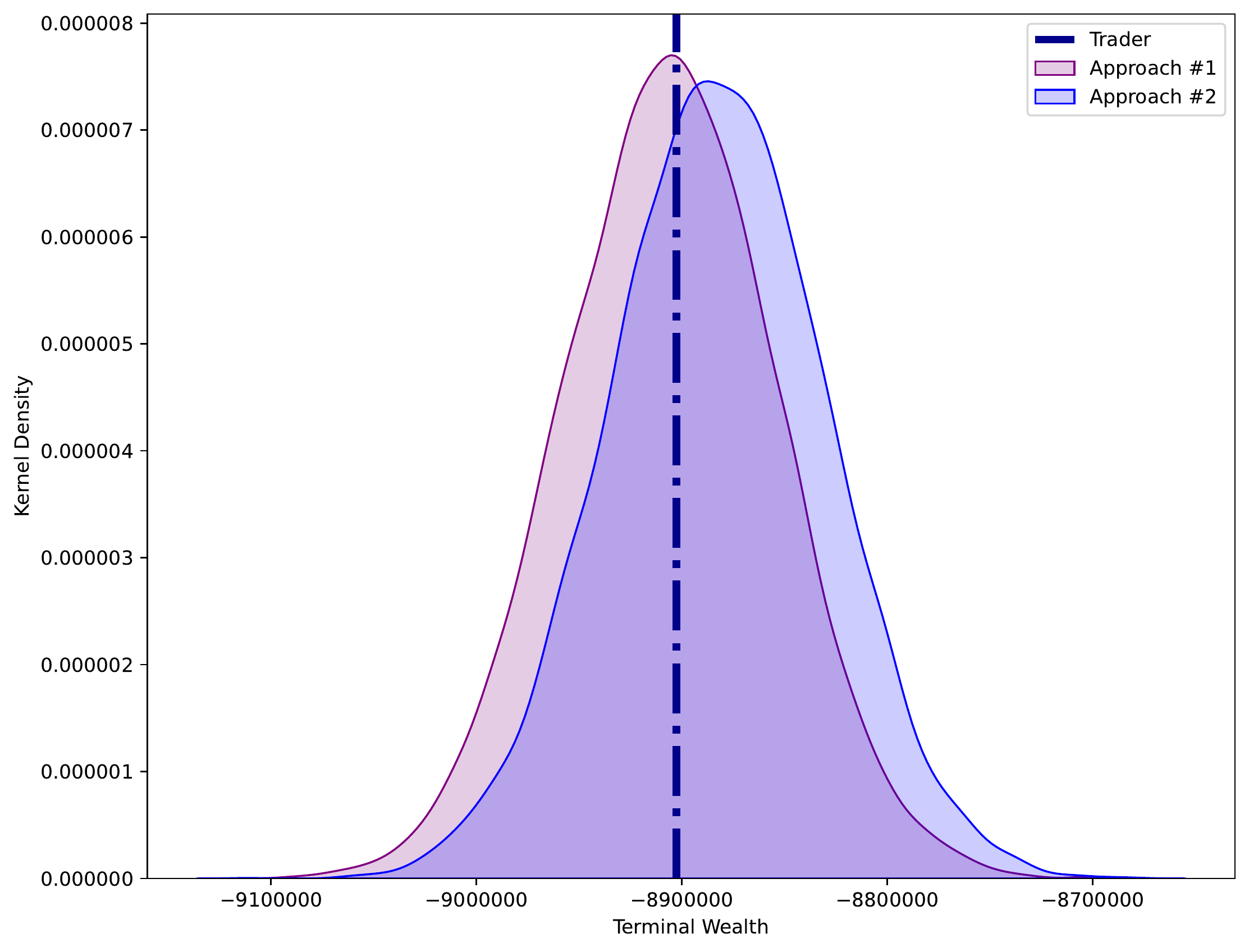}
\end{minipage}
\hspace{10mm}
\caption{Kernel density plots for terminal wealth $X_T$, on 08/27/2020. The black dashed vertical line represents the final wealth of the trader at the end of the day. The purple distribution presents the terminal wealth resulting from the Monte Carlo simulations of Approach $\#1$, while the distribution in blue results from Approach $\#2$.}
\label{fig:kernel_density}
\end{figure}

\vskip 6pt
Figure \ref{fi:inventory082720} shows that for both approaches, the inventories of the Monte Carlo simulations of the \emph{optimal} executions behave quite similarly to the actual time evolution of the inventory of Citadel on that day. On the other hand, the pattern observed in Figure \ref{fi:wealth082720} for the wealth deserves some comments. The first obvious remark is the apparent constraint on the terminal wealth of the band occupied by the scenarios. The form of the reward function does not constrain the terminal value of the wealth as the plot seems to indicate, and the actual distribution of the final wealth can be seen in Figure \ref{fig:kernel_density}. The second aspect of this figure is the raggedness of the scenarios, which reflects the presence of the new Brownian motion component and more closely emulates the actual wealth in red, as opposed to a differentiable function of time which would have had a smooth path.

\vskip 6pt

Figure \ref{fig:kernel_density} presents a comparison for the final wealth $X_T$ of the trader to kernel density plots of the final wealth obtained using Approaches $\#1$ and $\#2$. Approach $\#1$ ignores the price impact of the trades and directly uses the prices observed in the market. In doing so, it outputs slightly worse results on average when compared to the trader, only outperforming the trader on $47.62\%$ of the scenarios. Approach $\#2$, on the other hand, takes the permanent market impact that trading has on the price into consideration and, on average, implies better trading proceeds than the one obtained by the trader on this day, outperforming the trader on $64.99\%$ of the scenarios.

\begin{figure}
  \textbf{Inventory Evolution in Both Approaches}\par\medskip
\centerline{
\includegraphics[width=6.5cm,height=6.5cm]{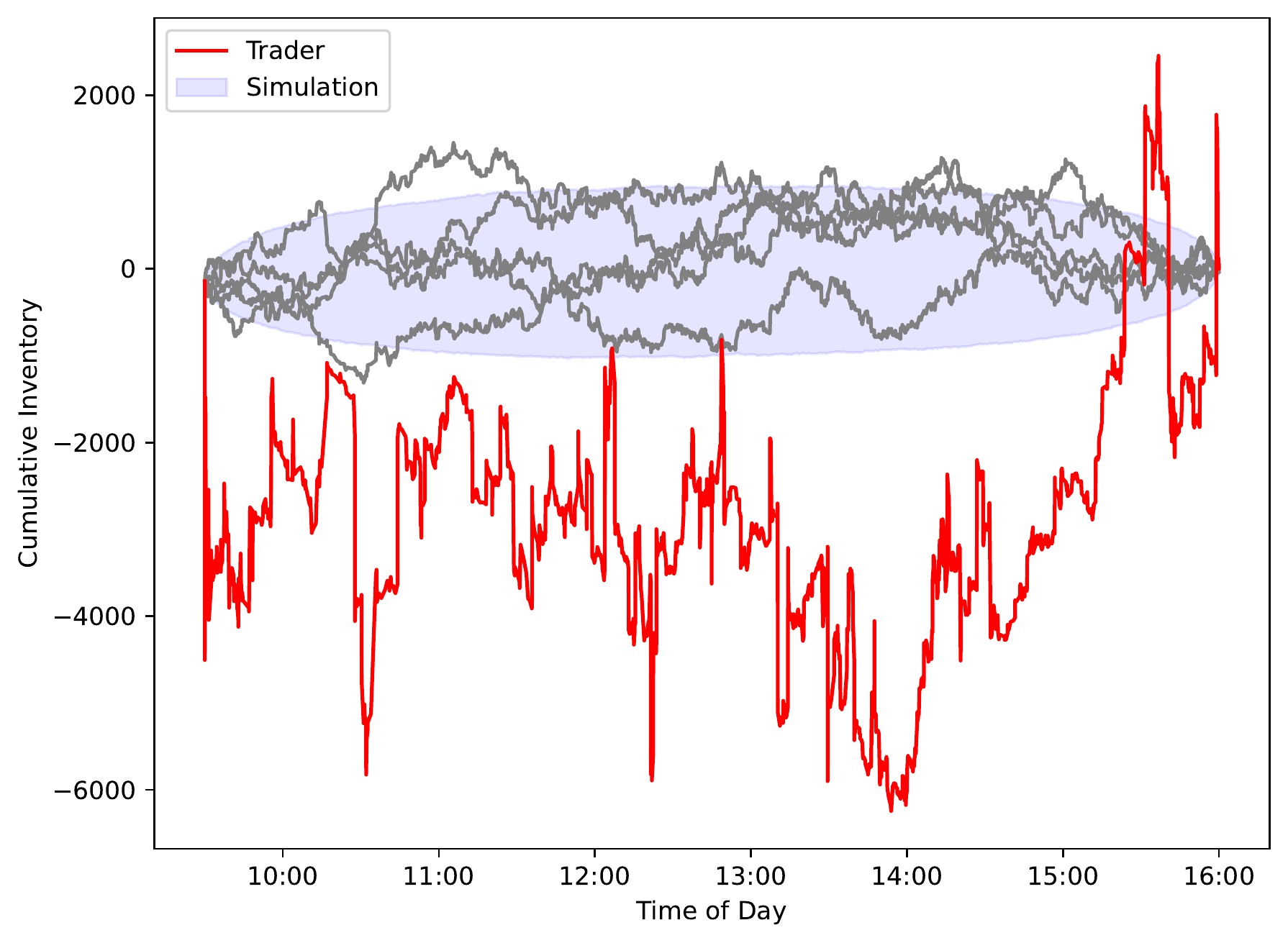}
\includegraphics[width=6.5cm,height=6.5cm]{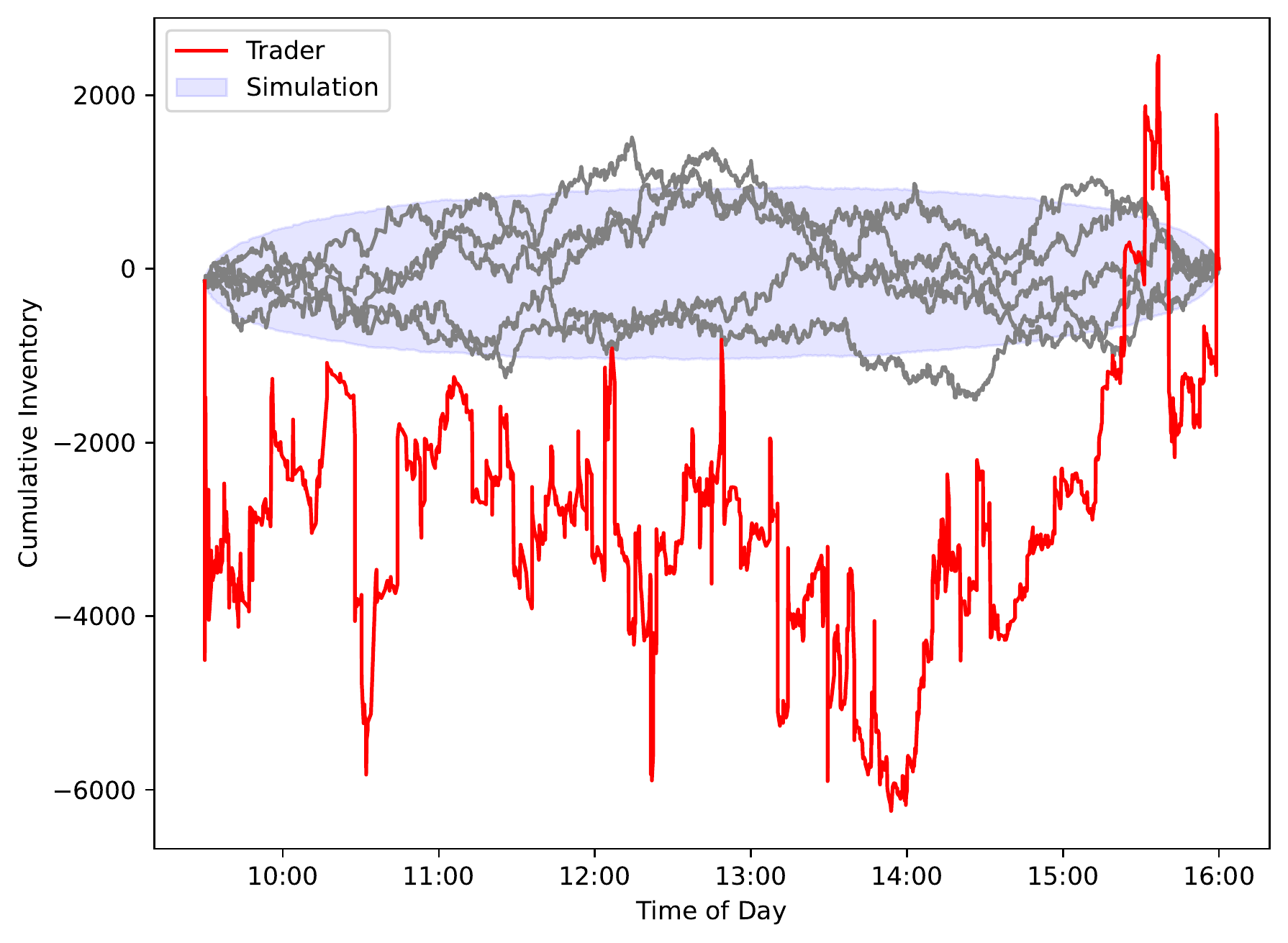}
}
\caption{Plot of the actual RY inventory (red curve) of Citadel on 08/27/2020 together with $5$ sample Monte Carlo scenarios of what the inventory would have been had the optimal rate of trading been used. The light blue band is bounded by the $5\%$-tile and the $90\%$-tile of $N_{sim}=10,000$ scenarios. The left pane gives the result of Approach \#1 and the plot in the right pane was produced following Approach \#2.} 
\label{fi:inventory032520}
\end{figure}

\begin{figure}
  \textbf{Wealth Evolution in Both Approaches}\par\medskip
\centerline{
\includegraphics[width=6.5cm,height=6.5cm]{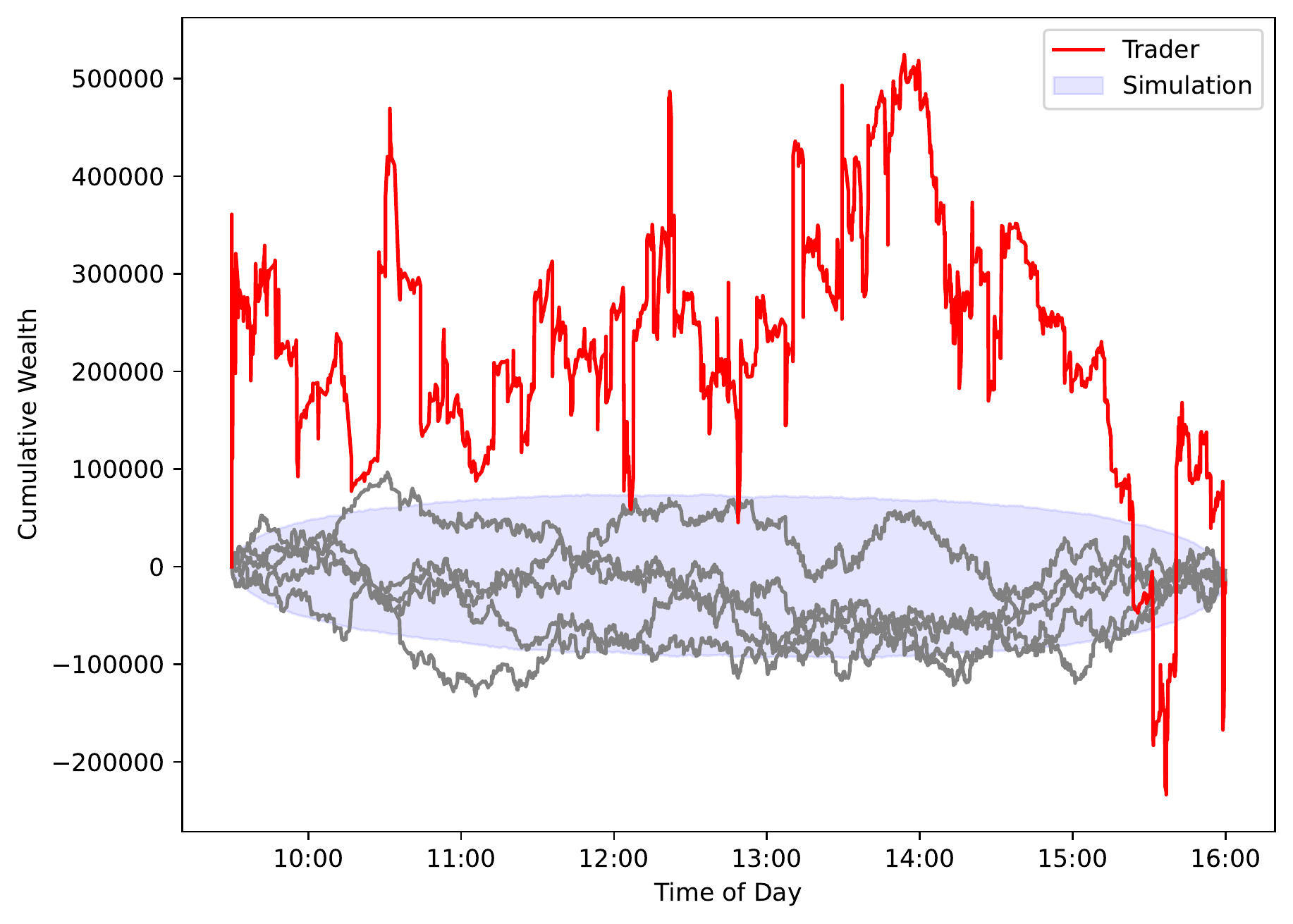}
\includegraphics[width=6.5cm,height=6.5cm]{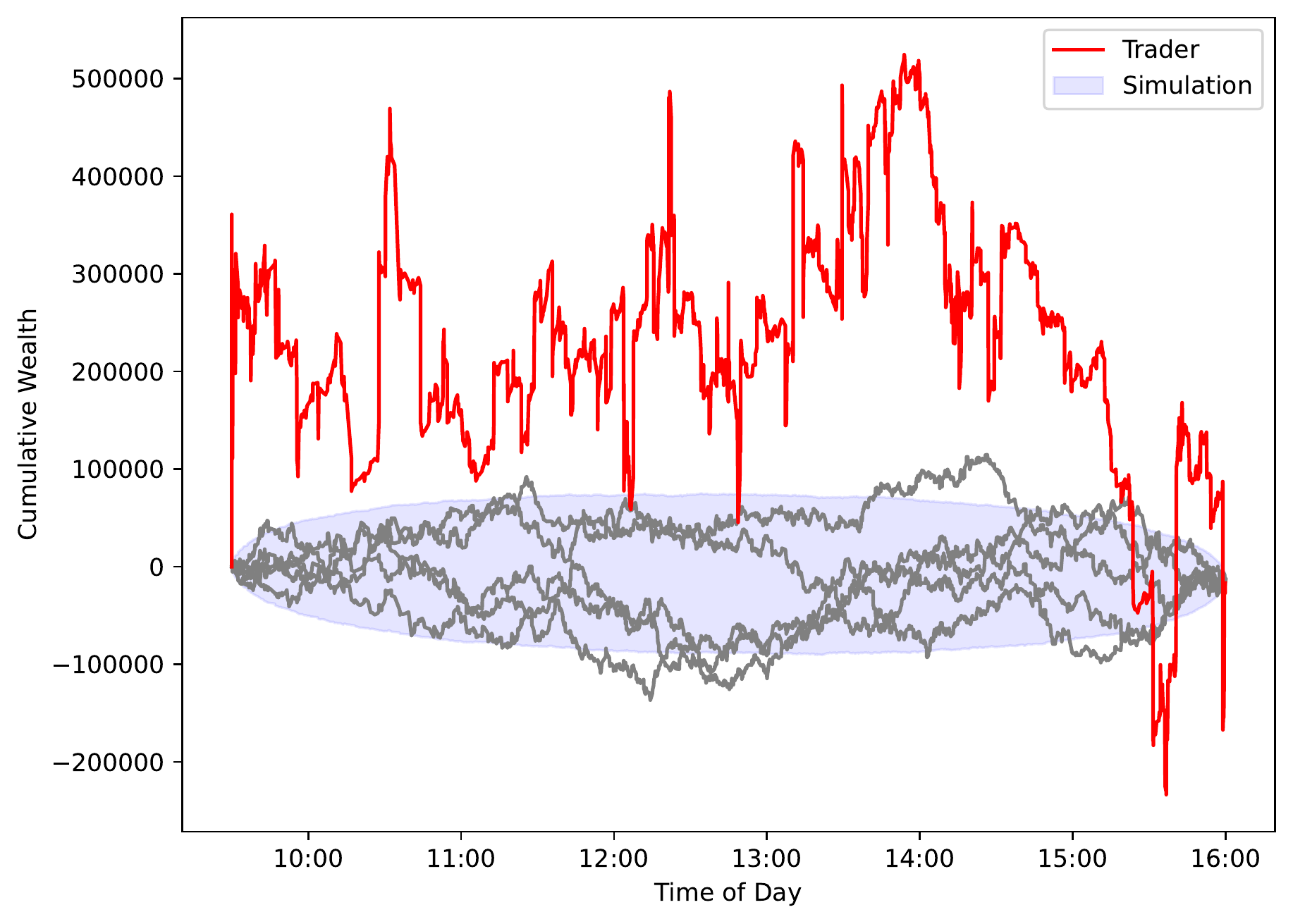}
}
\caption{Plot of the actual wealth (red curve) accumulated by Citadel by trading RY on 03/25/2020 together with $5$ sample Monte Carlo scenarios of what this wealth would have been had the optimal rate of trading been used. The light blue band is bounded by the $5\%$-tile and the $90\%$-tile of $N_{sim}=10,000$ scenarios. The left pane gives the result of Approach \#1 and the plot in the right pane was produced following Approach \#2.} 
\label{fi:wealth032520}
\end{figure}

\begin{figure}[H]
\textbf{Terminal Wealth Comparison, Kernel Density}\par\medskip
\centering
\begin{minipage}{.45\linewidth}
  \includegraphics[width=\linewidth]{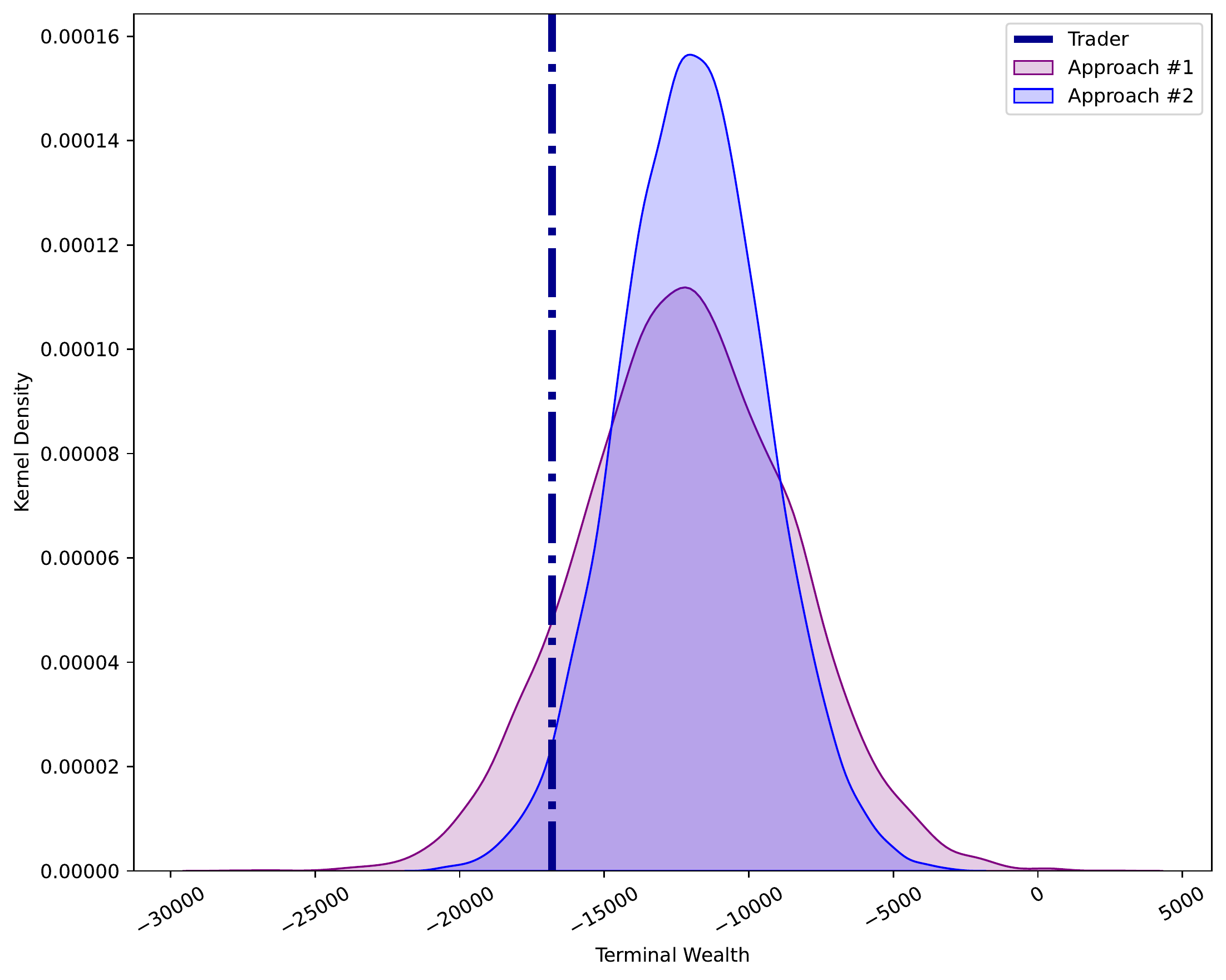}
\end{minipage}
\hspace{10mm}
\caption{Kernel density plots for terminal wealth $X_T$, on March $25^{th}$, 2020. The black dashed vertical line represents the final wealth of the trader at the end of the day. The purple distribution presents the terminal wealth resulting from the Monte Carlo simulations of Approach $\#1$, while the distribution in blue results from Approach $\#2$.}
\label{fig:kernel_density2}
\end{figure}

\vskip 12pt
Figure \ref{fi:inventory032520} and Figure \ref{fi:wealth032520} present the results of the same computations performed with the trading data of 03/25/2020. This day was chosen because the terminal inventory in RY shares of Citadel was basically the same as at the begining of the day, suggesting that Citadel was not trading on behalf of a client trying to aquire or liquidate a large position in the stock, but possibly trading to provide liquidity like a market maker. As expected, Figure \ref{fi:inventory032520} shows that the inventory scenarios follow the same pattern. While the comments we made on Figure \ref{fi:wealth082720} still apply, one more remark is in order. The actual terminal profit  and loss of Citadel for trading RY on that day is less than what should have been expected from the optimal strategy identified in the analysis of the theoretical model. Indeed, $90.13\%$ of the Monte Carlo scenarios ended up with a better result for Approach $\#1$, while $97.48\%$ of the Monte Carlo scenarios ended up with a better result than the trader for Approach $\#2$. The kernel density plot for the final wealth distribution can be seen in Figure \ref{fig:kernel_density2}.

\section{Conclusion}
\label{sec:4}

While the price process is well-known for being reasonably well modelled as a semi-martingale with a continuous martingale component, little is known about the nature of the inventory process and, consequently, about the wealth process of individual traders. In this work, we hope to have bridged that gap. 

Using high-frequency econometrics testing, we showed that both the inventory and the wealth dynamics should be modelled including Brownian motion components. We showed that, for active traders, this is true for both regularly re-sampled and asynchronous data. This is a powerful result because it sheds light on two processes that are usually modelled as differentiable functions of time, thus affecting dramatically the conclusions of most resulting analysis.

Furthermore, we apply this new knowledge about the presence of Brownian motion components in the inventory and wealth processes to a natural extension of a classic optimal execution problem. We show that the stochastic maximum principle can be used to solve the optimal control problem of the trader explicitly.

We then compare Monte Carlo simulated trajectories using the optimal control we obtained to the behavior of an actual trader. For the simulations, we use two different approaches. The first approach takes the price directly observed in the market, while the second considers the price impact resulting from the trading behavior. 

Finally, we compare these two approaches to the trader's behavior in two different situations. The first is when the trader is trying to execute a large buy order in the market. In this case, we find that the trader's behavior is very similar to what the model indicates in the first approach, but performs on average worse than when price impact is considered. The second situation is when the trader ends the trading day with virtually no change in their inventory. In this case, the trader's behavior is much less profitable than what the model indicates on average, indicating that a different setup than the optimal execution is more appropriate, for example, a market making model, to explain the actual behavior observed on the market.

\pagebreak

\bibliographystyle{chicago}

\pagebreak

\appendix 
\section{Parameters}

In this paper, we use ticker data for trades on equities traded on the Toronto Stock Exchange for the period that ranges from Jan/2008 until Oct/2020. The data is available for ten stocks. They represent a diverse range of industries, and all stocks present large daily traded volumes. We estimate the market impact parameters $\alpha$ and $\kappa$ as in \cite{LL1} for all the stocks simultaneously by adjusting for the average spread and average volume of the stock. The terminal penalty $A$ and the running penalty $\phi$ are taken to be 0.03 and 9.9e-7, respectively. The price and inventory volatilities are estimated from the asynchronous data observations in the trade file. The price volatility $\sigma_S$ is the average of the intraday standard deviation for the previous ten trading days, while the inventory volatility is the standard deviation of the inventory path of the trader on the given day. Tables \ref{tab:model_parameters} and \ref{tab:model_parameters2} summarize the parameters used in the paper:
\begin{table}[H]
\centering
\resizebox{.55\textwidth}{!}{
\begin{tabular}{lr}
 \hline
Terminal penalty, $\hat{A}$ & 0.03    \\ \hline
Running penalty, $\hat{\phi}$  & 9.9e-7  \\ \hline
Permanent Market Impact, $\hat{\alpha}$ & $0.22299 \cdot \frac{\text{avg. bin spread}}{\text{avg. bin volume}}\cdot \frac{1}{dt}$     \\\hline 
Temporary Market Impact $\hat{\kappa}$ & $0.07176 \cdot \frac{\text{avg. bin spread}}{\text{avg. bin volume}}\cdot \frac{1}{dt}$     \\ \hline
\end{tabular}
}
\caption{\small Model parameters, general.}
\label{tab:model_parameters}
\end{table}
 
\begin{table}[H]
\centering
\resizebox{.45\textwidth}{!}{
\begin{tabular}{lr}
 \hline
Permanent Market Impact, $\hat{\alpha}$, for RY & 1.27e-06 \\\hline 
Temporary Market Impact $\hat{\kappa}$ for RY &  4.07e-07 \\ \hline
Price volatility, $\hat{\sigma_S}$ on 08/27/2020 & 0.21\\ \hline
Price volatility, $\hat{\sigma_S}$ on 03/25/2020 & 1.49\\ \hline
Inventory volatility, $\hat{\sigma_Q}$ on 08/27/2020  & 25707.43\\ \hline
Inventory volatility, $\hat{\sigma_Q}$ on 03/25/2020 & 1449.32\\ \hline
\end{tabular}
}
\caption{\small Model parameters, specific cases discussed in the paper.}
\label{tab:model_parameters2}
\end{table}

\pagebreak

\appendix

\begin{table}[]
\centering
\begin{tabular}{cl}
\hline
\textbf{TSX TRADING \#} & \multicolumn{1}{c}{\textbf{PARTICIPATING ORGANIZATIONS}}      \\ \hline
1                       & Anonymous \\
2                       & RBC Capital Markets  \\
3                       & ATB Capital Markets Inc. \\
4                       & Cantor Fitzgerald Canada Corporation  \\
5                       & Citadel Securities Canada ULC \\
6                       & INFOR Financial Inc.  \\
7                       & TD Securities Inc.  \\
8                       & Eight Capital  \\
9                       & BMO Nesbitt Burns Inc. \\
11                      & Macquarie Capital Markets Canada Ltd. \\
12                      & Bloom Burton Securities Inc.  \\
13                      & Instinet Canada Ltd.  \\
14                      & Virtu ITG Canada Corp. \\
15                      & UBS Securities Canada Inc./UBS Valeurs Mobilieres Canada Inc. \\
16                      & Paradigm Capital Inc.\\
17                      & Integral Wealth Securities Limited  \\
18                      & Echelon Wealth Partners Inc.   \\
19                      & Desjardins Securities Inc.  \\
21                      & Pavilion Global Markets Ltd.    \\
22                      & Fidelity Clearing Canada ULC    \\
23                      & State Street Global Markets Canada Inc.   \\
24                      & Clarus Securities Inc. \\
25                      & Odlum Brown Ltd.  \\
28                      & CI Investment Services Inc.   \\
33                      & Canaccord Genuity Corp. \\
34                      & Maison Placements Canada Inc. \\
35                      & Friedberg Mercantile Group  \\
36                      & W.D. Latimer Co. Ltd. \\
38                      & Liquidnet Canada Inc.  \\
39                      & Merrill Lynch Canada Inc.  \\
43                      & Caldwell Securities Ltd. \\
48                      & Laurentian Bank Securities Inc.  \\
53                      & Morgan Stanley Canada Ltd.  \\
55                      & Velocity Trade Capital Ltd.  \\
56                      & Edward Jones \\
57                      & Interactive Brokers Canada Inc.   \\
59                      & PI Financial Corp.  \\
62                      & Haywood Securities Inc.  \\
65                      & Goldman Sachs Canada Inc. \\
68                      & Leede Jones Gable Inc.  \\
70                      & Manulife Securities Incorporated \\

\end{tabular}
\end{table}

\begin{table}[]
\centering
\begin{tabular}{cl}
\hline
\textbf{TSX TRADING \#} & \multicolumn{1}{c}{\textbf{PARTICIPATING ORGANIZATIONS}}      \\ \hline
72                      & Credit Suisse Securities (Canada), Inc. \\
73                      & Cormark Securities Inc./Valeurs Mobilieres Cormark Inc.    \\
74                      & RF Securities Clearing LP (RF Clearing)  \\
76                      & Industrial Alliance Securities Inc.  \\
77                      & Peters \& Co. Ltd.   \\
79                      & CIBC World Markets  Inc.  \\
80                      & National Bank Financial Inc.  \\
81                      & HSBC Securities (Canada) Inc. \\
82                      & Stifel Nicolaus Canada Inc  \\
83                      & Mackie Research Capital Corp. \\
84                      & Independent Trading Group  \\
85                      & Scotia Capital Inc.\\
86                      & Pictet Canada L.P.  \\
87                      & Beacon Securities Ltd. \\
88                      & Credential Qtrade Securities Inc.  \\
89                      & Raymond James Ltd. \\
90                      & Barclays Capital Canada Inc. \\
91                      & JonesTrading Canada Inc. \\
92                      & Pollitt \& Co. Inc.\\
94                      & Hampton Securities Ltd.  \\
97                      & M Partners Inc. \\
101                     & Société Générale Capital Canada Inc. \\
123                     & Citigroup Global Markets Canada  \\
124                     & Questrade Inc.  \\
143                     & Pershing Securities Canada Ltd.   \\
150                     & Generation IACP Inc \\
200                     & Acumen Capital Finance Partners Ltd.   \\
201                     & Wellington-Altus Private Wealth Inc.   \\
203                     & Assante Capital Management Ltd.   \\
208                     & Kingwest and Company \\
212                     & Virtu Financial Canada ULC   \\
222                     & JP Morgan Securities Canada Inc. \\
234                     & Clearpool Execution Services (Canada) Limited     
\end{tabular}
\end{table}

\end{document}